\documentclass[ modern]{aastex631}

\hypersetup{linkcolor=red,citecolor=blue,filecolor=cyan,urlcolor=magenta}
\usepackage{url}
\usepackage{etex}
\usepackage{amsmath,amssymb,latexsym} 
\usepackage{xcolor, fontawesome} 
\usepackage{graphicx}
\usepackage{lipsum}
\usepackage{graphicx}
\usepackage{appendix}
\usepackage{pgfplots}
\pgfplotsset{compat=1.3}
\usepackage{color,soul}
\usepackage{needspace}

\begin{document}

\title{Hydrodynamic Predictions for the Next Outburst of T Coronae Borealis: It will be the Brightest Classical or Recurrent Nova Ever Observed in 
X-rays \footnote{RMW took part in the observations and early simulations for this paper and we are grateful for his help.}}

\author[0000-0002-1359-6312]{S. Starrfield}
\affiliation{Earth and Space Exploration, Arizona State University, P.O. Box 876004, Tempe, AZ, 85287-6004, USA
starrfield@asu.edu}

\author[0000-0002-7978-6570]{M. Bose}
\affiliation{Earth and Space Exploration, Arizona State University, P.O. Box 876004, Tempe, AZ, 85287-6004, USA}
\affiliation{Center for Isotope Analysis (CIA), Arizona State University, Tempe, AZ, 85287-6004, USA}

\author[0000-0001-6567-627X]{C. E. Woodward}
\affiliation{MN Institute for Astrophysics, 116 Church Street, SE University of Minnesota, Minneapolis, MN 55455, USA }

\author[0000-0003-2381-0412]{C. Iliadis}
\affiliation{Department of Physics \& Astronomy, University of North Carolina, Chapel Hill, NC 27599-3255, USA}
\affiliation{Triangle Universities Nuclear Laboratory, Durham, NC 27708-0308, USA}

\author[0000-0002-9481-9126]{W. R. Hix}
\affiliation{Physics Division, Oak Ridge National Laboratory, Oak Ridge TN, 37831-6354, USA}
\affiliation{Department of Physics and Astronomy, University of Tennessee, Knoxville, TN 37996, USA }

\author[0000-0002-3142-8953]{A. Evans}
\affiliation{Astrophysics Group, Keele University, Keele, Staffordshire, ST5 5BG, UK}

\author[0000-0003-4615-8009]{G. Shaw}
\affiliation{Department of Astronomy and Astrophysics, Tata Institute of Fundamental Research, 
Homi Bhabha Road, Mumbai 4000005, India}

\author[0000-0002-9670-4824]{D. P. K. Banerjee}
\affiliation{Physical Research Laboratory, Navrangpura, Ahmedabad, Gujarat 380009, India}

\author[0000-0003-2196-9091]{T. Liimets}
\affiliation{Tartu Observatory, University of Tartu, Observatooriumi 1, T\~oravere 61602, Estonia}

\author[0000-0001-5624-2613]{K. L. Page}
\affiliation{School of Physics and Astronomy, University of Leicester, Leicester, LE1 7RH, UK}

\author[0000-0003-2824-3875]{T. R. Geballe}
\affiliation {Gemini Observatory/NSF's NOIRLab, 670 N. Aohoku Pl., Hilo, HI 96720, USA}

\author[0000-0002-0551-046X]{I. Ilyin}
\affiliation {Leibniz-Institut fur Astrophysik Potsdam: Potsdam, Brandenburg, DE}

\author[0009-0001-5755-8297]{I. Perron}
\affiliation{MN Institute for Astrophysics, 116 Church Street, SE University of Minnesota, Minneapolis, MN 55455, USA }

\author[0000-0003-1892-2751]{R. M. Wagner}
\affiliation{Large Binocular Telescope Observatory, Tucson, AZ 85721, USA}
\affiliation{Department of Astronomy, Ohio State University, Columbus, OH 43210, USA}
\affiliation{deceased, September 2, 2023}

\newcommand{\kms}{km\,s$^{-1}$}
\def\lesssim{\mathrel{\hbox{\rlap{\hbox{\lower4pt\hbox{$\sim$}}}\hbox{$<$}}}}
\def\gtrsim{\mathrel{\hbox{\rlap{\hbox{\lower4pt\hbox{$\sim$}}}\hbox{$>$}}}}
\def\apj{$Astrophys.\ J.$}
\def\apjl{$Astrophys.\ J.$}
\def\aj{$Astron.\ J.$}
\def\aap{$Astron.\ Astrophys.$}
\def\mnras{$Mon.\ Not.\ R.\ Astron.\ Soc.$}
\def\pasj{$Publ.\ Astron.\ Soc.\ Jpn$}
\def\iaucirc{$IAU Circ.$}
\def\n{\footnotemark}
\def\etal{{et al.}~}
\def\vla{{\it VLA}}
\def\iue{{\it IUE}}
\def\hst{{\it HST}}
\def\gro{{\it GRO}}
\def\batse{{\it BATSE}}
\def\rosat{{\it ROSAT}}
\def\ginga{{\it GINGA}}
\def\xte{{\it RXTE}}
\def\degK{$^{\rm o}$K}
\def\kms{\ifmmode {\rm km\ s}^{-1} \else km s$^{-1}$\fi}
\def\Msun{\ifmmode {\rm M}_{\odot} \else M$_{\odot}$\fi}
\def\Rsun{\ifmmode {\rm R}_{\odot} \else R$_{\odot}$\fi}
\def\Lsun{\ifmmode {\rm L}_{\odot}~ \else L$_{\odot}$\fi~}
\def\qo{\ifmmode q_{\rm o} \else $q_{\rm o}$\fi}
\def\Ho{\ifmmode H_{\rm o} \else $H_{\rm o}$\fi}
\def\Av{\ifmmode A_{\rm V} \else $A_{\rm V}$\fi}
\def\Mv{\ifmmode M_{\rm V} \else $M_{\rm V}$\fi}
\def\mv{\ifmmode m_{\rm V} \else $m_{\rm V}$\fi}
\def\Ha{\ifmmode {\rm H}\alpha \else H$\alpha$\fi}
\def\Hb{\ifmmode {\rm H}\beta \else H$\beta$\fi}
\def\Hg{\ifmmode {\rm H}\gamma \else H$\gamma$\fi}
\def\Hd{\ifmmode {\rm H}\delta \else H$\delta$\fi}
\def\Hep{\ifmmode {\rm H}\epsilon \else H$\epsilon$\fi}
\def\Lya{\ifmmode {\rm Ly}\alpha \else Ly$\alpha$\fi}
\def\Lyb{\ifmmode {\rm Ly}\beta \else Ly$\beta$\fi}
\def\hi{\ion{H}{1}}
\def\hii{\ion{H}{2}}
\def\hei{\ion{He}{1}}
\def\heii{\ion{He}{2}}
\def\ci{\ion{C}{1}}
\def\cii{\ion{C}{2}}
\def\ciii{\ion{C}{3}}
\def\civ{\ion{C}{4}}
\def\ni{\ion{N}{1}}
\def\nii{\ion{N}{2}}
\def\niii{\ion{N}{3}}
\def\niv{\ion{N}{4}}
\def\nv{\ion{N}{5}}
\def\oi{\ion{O}{1}}
\def\oii{\ion{O}{2}}
\def\oiii{\ion{O}{3}}
\def\oiv{\ion{O}{4}}
\def\ov{\ion{O}{5}}
\def\ovi{\ion{O}{6}}
\def\neiii{\ion{Ne}{3}}
\def\neiv{\ion{Ne}{4}}
\def\nev{\ion{Ne}{5}}
\def\mgi{\ion{Mg}{1}}
\def\mgii{\ion{Mg}{2}}
\def\siiv{\ion{Si}{4}}
\def\si{\ion{S}{1}}
\def\sii{\ion{S}{2}}
\def\siii{\ion{S}{3}}
\def\siv{\ion{S}{4}}
\def\cai{\ion{Ca}{1}}
\def\caii{\ion{Ca}{2}}
\def\fei{\ion{Fe}{1}}
\def\feii{\ion{Fe}{2}}
\def\feiii{\ion{Fe}{3}}
\def\fevii{\ion{Fe}{7}}
\def\fex{\ion{Fe}{10}}
\def\fexi{\ion{Fe}{11}}
\def \icarus  {Icarus, }
\def \vistas  {Vistas Astron., }
\def  \ergcms      {\hbox{erg cm$^{-2}$ s$^{-1}$}}   
\def  \erggm      {\hbox{erg gm$^{-1}$ s$^{-1}$}}   
\def  \Nh          {$ N_{\rm H} $}           
\def  \ctpers         {\hbox{counts s$^{-1}$}}  
\def  \adupers         {\hbox{ADU s$^{-1}$}}    
\def  \Fn          {$ F_{\nu} $}             
\def  \Fl          {$ F_{\lambda} $}         
\def  \Ln          {$ L_{\nu} $}
\def  \LEdd        {$ L_{\rm Edd} $}         
\def  \mdot        {$ \dot{m} $}             
\def  \Mdot        {$ \dot{M} ~ $}             
\def  \Myr         {\Msun yr$^{-1}$}         
\def  \Te          {$ T_{\rm e} $}           
\def  \Teff        {$ T_{\rm eff} $}         
\def  \Ne          {$ N_{\rm e} $}           
\def  \ergpers        {\hbox{erg s$^{-1}$}}     
\def  \ergscm2Hz    {\hbox{erg s$^{-1}$ cm$^{-2}$ Hz$^{-1}$}} 
\def  \ergscm2A     {\hbox{erg s$^{-1}$ cm$^{-2}$ \AA$^{-1}$}} 
\def  \percc          {\hbox{cm$^{-3}$}}        
\def  \percm2         {cm$^{-2}$}               
\def  \percm3         {cm$^{-3}$}               
\def  \cmpers         {\hbox{cm s$^{-1}$}}      
\def  \micron      {$\mu$m}                  
\def  \vs          {{\it vs.} }
\def \kms {km\,s$^{-1}$}
\def \mtrs {m\,s$^{-1}$}
\def\lesssim{\mathrel{\hbox{\rlap{\hbox{\lower4pt\hbox{$\sim$}}}\hbox{$<$}}}}
\def\gtrsim{\mathrel{\hbox{\rlap{\hbox{\lower4pt\hbox{$\sim$}}}\hbox{$>$}}}}
\def\apj{$Astrophys.\ J.$}
\def\apjl{$Astrophys.\ J.$}
\def\aj{$Astron.\ J.$}
\def\aap{$Astron.\ Astrophys.$}
\def\mnras{$Mon.\ Not.\ R.\ Astron.\ Soc.$}
\def\pasj{$Publ.\ Astron.\ Soc.\ Jpn$}
\def\iaucirc{$IAU Circ.$}
\def\7Be{$^7$Be}
\def\7Li{$^7$Li}

\newcommand{\ltsimeq}{\raisebox{-0.6ex}{$\,\stackrel
        {\raisebox{-.2ex}{$\textstyle <$}}{\sim}\,$}}
\newcommand{\gtsimeq}{\raisebox{-0.6ex}{$\,\stackrel
        {\raisebox{-.2ex}{$\textstyle >$}}{\sim}\,$}}
\newcommand{\prpsimeq}{\raisebox{-0.6ex}{$\,\stackrel
        {\raisebox{-.2ex}{$\propto$}}{\sim}\,$}}



\begin{abstract}
T Coronae Borealis (TCrB) is a recurrent nova (RN) with recorded outbursts in 1866, and 1946 {and
possible outbursts in 1217 and 1787}.
 It is predicted to explode again in 2025 or 2026 based on multiple observational studies. 
 The system consists of a massive (M$_{wd}$ $\gtrsim$ 1.35 \Msun) white dwarf (WD) and a red giant (M3-M4 III). 
 We have performed 1-D hydrodynamic simulations with NOVA to predict the behavior of the next outburst.
These simulations consist of a range of mass accretion rates onto $\sim$1.35 M$_\odot$ WDs, 
designed to bound the conditions necessary to achieve ignition of an explosion after an $\approx$80 year inter-outburst period. 
We have used both carbon-oxygen and oxygen-neon initial compositions, in order to include the possible ejecta abundances 
to be measured in the observations of the next outburst. As the WD in the TCrB system is observed to be massive, 
theoretical predictions reported here imply that the WD is growing in mass as a consequence of the TNR.  
Therefore, the secular evolution of the WD may allow it to approach
 the Chandrasekhar limit and either explode as a Type Ia supernova or undergo accretion induced collapse, 
 depending on its underlying composition. We have followed the evolution of just the WD, after removing the ejected matter from the surface layers. Our intent is to illuminate the mystery of the unique, second, maximum in the two well observed outbursts and we have found conditions that
bracket the predictions.

 \end{abstract}


\section{T Coronae Borealis: What we know so far}
\label{intro}

Recurrent novae (RNe) such as T Coronae Borealis (TCrB) are members of the class of Cataclysmic Variables (CVs) which consist of
a white dwarf (WD) orbiting a secondary star.  In Classical Novae (CNe) the secondary is near or on the Main Sequence and the orbital
periods are typically on the order of hours. These systems are transferring matter from the secondary onto the WD and eventually a 
thermonuclear runaway (TNR) occurs that explosively ejects both the accreted matter and matter mixed up from the outer 
layers of the WD. The consequences are observed as a nova outburst.  In a few cases the secondaries are massive or evolved and
the orbital periods range from a few days to months and the recurring outbursts occur on a human timescale. These are designated
as RNe.  Reviews of the outbursts of both CNe and RNe can be found in
\citet{anupama_2008_aa, darnley_2020_aa, darnley_2021_aa,dellavalle_2020_aa, chomiuk_2021_aa, munari_2024_aa}.

TCrB is a Symbiotic Recurrent Nova (SyRN).  It
consists of a massive WD orbiting a red giant (M3-M4 III) with an orbital period of $\sim$227.6 days \citep{anupama_1999_aa,
anupama_2008_aa}. TCrB had possible outbursts  in 1217 and 1787 \citep{schaefer_2023_ab}, and recorded outbursts in 1866 and 1946. Based on its current behavior,  it is likely that it will explode again in the next year or two \citep{munari_2016_aa, munari_2023_aa, munari_2023_ab, Schaefer_2023_aa}.

TCrB reached a peak visual magnitude of $\sim$2 in 1946  \citep{schaefer_2022_aa} and its outburst was well observed (for the day)  both photometrically and spectroscopically
\citep{mclaughlin_1946_aa, herbig_1946_aa, pettit_1946_aa, morgan_1947_aa, sanford_1947_aa, sanford_1949_aa, payne-gaposchkin_1964_aa}.  Its distance is $\lesssim$ 1 kpc \citep{schaefer_2018_aa, schaefer_2022_aa} but the amount of mass ejected in radioactive nuclei by a RN is far less than in a CN, suggesting that TCrB is unlikely to be detected in nuclear emitting $\gamma$-rays during its coming outburst (see Section \ref{nucleo}).

The first  determinations of the mass of the WD in the system suggested that it exceeded the Chandrasekhar limit \citep{kraft_1958_aa, kraft_1964_aa, kenyon_1986_aa}, but more recent studies have reanalyzed their data and added more radial velocity measurements so that
the current best value of the WD mass is $\sim$1.35 M$_\odot$ \citep{selvelli_1992_aa, shahbaz_1997_aa, belczynski_1998_aa, stanishev_2004_aa, kennea_2009_aa, shara_2018_ab, selvelli_2019_aa}.  Assuming it takes $\sim10^{-6}$~M$_\odot$ to initiate a TNR on such a massive WD (see Section \ref{results}), an 80 yr inter-outburst time implies an average \.M $\sim10^{-8}$ \Myr.

\begin{figure}[htb!]
\includegraphics[width=1.0\textwidth]{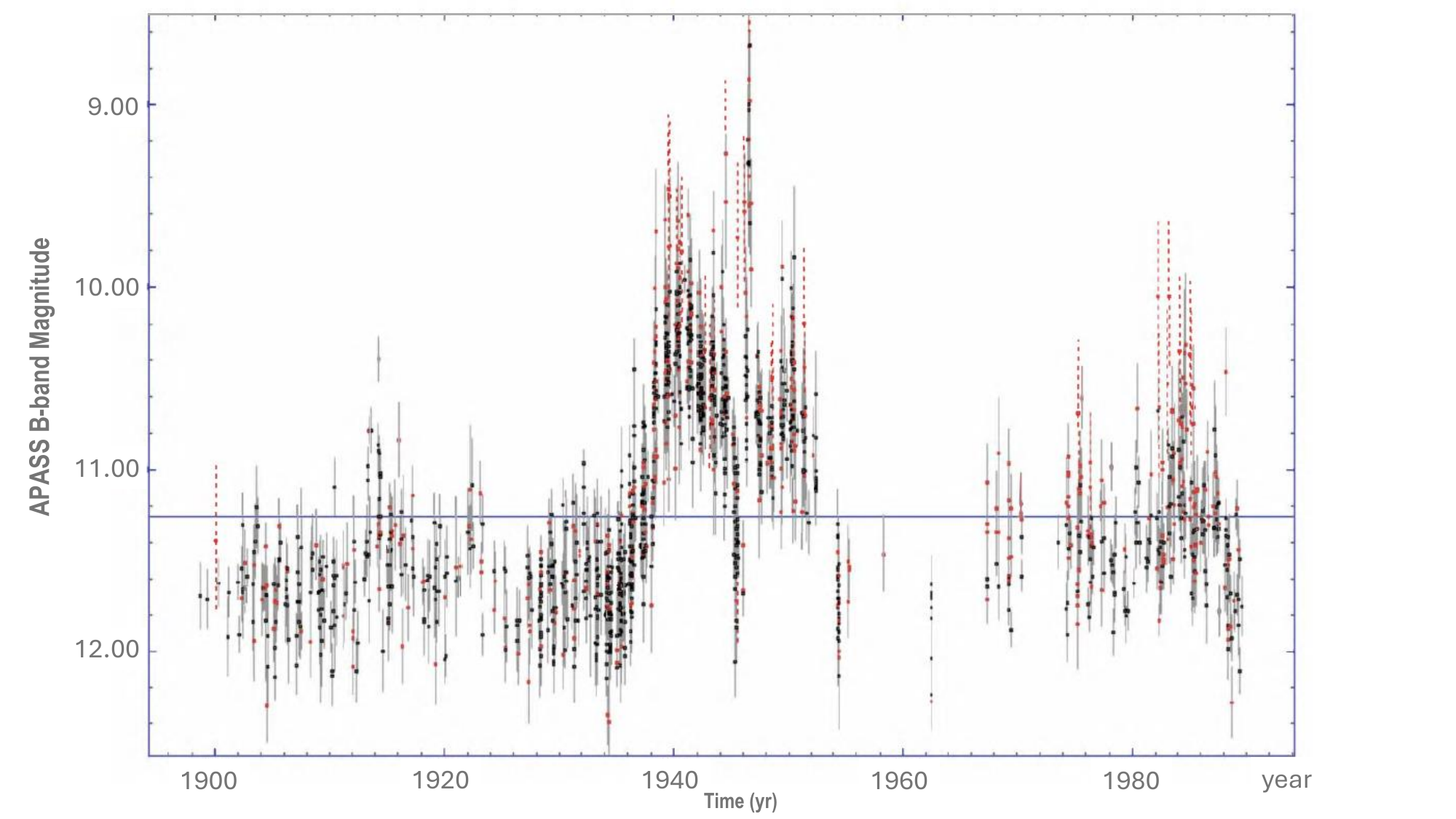}
\caption{The DASCH \citep{grindlay_2012_aa, dasch_2024_aa} light curve of TCrB from 1900 to 1990.  The y-axis makes use of the
AAVSO Photometric All-Sky Survey (APASS)  B-band magnitudes. The x-axis is the time in years.  There are periods of enhanced brightness probably designating times of increased \.M. The outburst is clearly visible in 1946. }
\label{dasch}
\end{figure}

A distinguishing feature that separates TCrB from the other SyRNe (except for RT 
Cru) is that it exhibits hard X-ray emission \citep{kennea_2009_aa}. 
When \citet{kennea_2009_aa} compared TCrB to other types of CVs that also showed hard X-ray emission (notably magnetic CVs), they concluded that 
they were observing accretion onto a massive WD.  

However, the UV luminosity of TCrB, measured over some years with the International Ultraviolet 
Explorer (IUE) satellite, far exceeds that in the X-ray regime. 
\citet{selvelli_1992_aa} and \citet{kennea_2009_aa} attribute this difference to an optically thick boundary layer. 
The luminosity reported by \citet{selvelli_1992_aa} was determined assuming a distance to TCrB of 1.3 kpc.  
Reducing the distance to a value of $\sim$900 pc \citep{schaefer_2018_aa, schaefer_2022_aa}, reduces their reported luminosity 
by about a factor of 2 to $\sim 7.7 \times 10^{34}$ \ergpers ($\sim$ 20 L$_{\odot}$).  
Assuming this luminosity is produced by accretion onto the WD (L = 0.5$\times$GM\.M/R where M and R are the mass and radius of the WD and \.M is the mass accretion rate),  the radii for 1.35M$_\odot$ WDs used in this paper, and listed in Table 1, yield an accretion rate  that is too low for the WD to accumulate sufficient mass to initiate a TNR in 80 years.

The probable solution to this enigma is that the rate of mass accretion onto
the WD is variable and most of the material is accreted a few years prior and a few years after the actual outburst \citep{munari_2016_aa, luna_2020_aa, munari_2023_aa}. A detailed discussion of this behavior can be found in \citet{luna_2020_aa}.  They use a distance to TCrB of 806 pc, however, which is smaller than that given by  \citet{schaefer_2022_aa}.,  Figure \ref{dasch} shows the long term light curve of TCrB obtained from the Harvard Plates 
by the DASCH project \citep{dasch_2024_aa}.  Clearly, there are periods, outside the actual outburst in 1946,  when the system is nearly 2 magnitudes brighter than ``normal'' and it is likely during those periods when most of the mass is accreted as discussed by \citet{munari_2016_aa}, \citet{luna_2020_aa}, and \citet{munari_2023_aa}.

An apparently unique aspect of the TCrB light curve is the second maximum that occurs
about 100 days after the initial outburst. A plot showing both the initial outburst and the secondary maximum can be
found in \citet[][Figure 1]{munari_2023_aa}.  The cause of the second maximum is still unexplained, but \citet{munari_2023_aa} proposed
that it is caused by the evolution of the WD after the explosion.  \citet{munari_2023_aa}
has examined the effects of the radiation from a cooling WD ($\sim 2\times 10^5$ K to $\sim 10^5$ K over a 150 day interval)
reflecting off the red giant companion.  By including the changing aspect of the WD plus red giant in the system, \citet{munari_2023_aa}
obtains a light curve that closely resembles the actual secondary maximum of TCrB.  We test this theory by following the evolution of just the WD
after the ejected matter is removed from the simulation and report those results in Section \ref{remnant}.

In Section \ref{novacode}, we provide a brief discussion of our 1D hydrodynamic computer code, NOVA. We follow that with the results of our simulations in Section \ref{results}, a summary of the nucleosynthesis in Section \ref{nucleo}, the evolution of the WD to
quiescence in Section \ref{remnant}, a discussion in Section \ref{discuss}, and end with the conclusions in Section \ref{conclude}.


\section{The Hydrodynamic Code}
\label{novacode}

We use NOVA, a  one-dimensional (1-D), fully implicit, hydrodynamic, computer code \citep[][]{kutter_1972_aa, sparks_1972_aa, kutter_1974_aa, kutter_1980_aa}. The most recent descriptions of NOVA can be found in 
\citet[][and references therein]{starrfield_2009_aa, starrfield_2020_aa, starrfield_2024_aa}. The major change is to the initial WD structure.
Previously, CNe modeling with NOVA has used complete WDs (the WD mass zones extend to the center), and used the Equations
of State (EOS) and opacities included in NOVA to  converge to a structure initially in hydrostatic equilibrium.

Recently,  new calculations of the structure of massive WDs (M$_{wd}$ $\gtrsim$ 1.35 \Msun) have appeared both with an improved EOS and now including General Relativity (GR).  
They are \citet[][Table 1]{althaus_2022_ab} for oxygen-neon (ONe) WDs and \citet[][Table 1]{althaus_2023_aa} for carbon-oxygen (CO) WDs.  They describe in great detail the improvements to the structures of the WDs and those are not repeated here.
The 1.35 M$_\odot$ ONe radii in \citet{althaus_2022_ab} (GR: 1542.51 km and non-GR: 1829.29 km) are significantly smaller than those computed with NOVA as complete WDs.  

Therefore, in order to use their radii, 
we consider only WD envelopes. We arbitrarily choose an envelope mass of 0.13 M$_\odot$ (chosen to be far bigger than the accreted mass), 
and then iterate on the inner radius until the WD outer radius nearly agrees (1522 km as opposed to 1542 km and 1827 km instead of 1829 km) with the radius published by \citet{althaus_2022_ab} in their ONe paper.  The inner and outer radii (the latter labeled: White Dwarf Radius) for each of the two ONe WDs are given in Table \ref{tcrbtable1}. 

However, the 1.35 M$_\odot$ CO WD radius computed by NOVA using Newtonian gravity (2166 km; see Table \ref{tcrbtable1}) lies in between the non-GR (2184 km) and GR (2027 km) radii published by \citet{althaus_2023_aa} so the NOVA radius, and a complete WD, was used in this study. 

While our two ONe radii and the CO radius agree only approximately with the published values in \citet{althaus_2022_ab, althaus_2023_aa}, upon examining their two tables it is clear that small changes in WD mass (Column 1 in Table 1 in both \citet{althaus_2022_ab} and \citet{althaus_2023_aa}) result in sufficient changes in the radius to encompass the radii that we use.  For example, increasing the CO WD mass to 1.382~M$_\odot$ reduces the WD  radii reported in \citet{althaus_2022_ab}  to 1639 km (non-GR) and 1066 km (GR).  Thus, we are justified to use the same set of compositions for all three radii.

We use five different compositions for the accreted matter in this study. The first is a solar mixture using the abundances in \citet{lodders_2003_aa}.   We continue
to use that solar composition to enable a comparison of the results in this study to earlier work.  The first two of the four mixed compositions are
25\% ONe WD matter and 75\% solar matter  \citep{lodders_2003_aa} (ONe 25/75) and  50\% ONe WD matter and 
50\% solar matter \citep{lodders_2003_aa} (ONe 50/50) which allows us to compare our results to \citet{hernanz_1996_aa}, \citet{jose_1998_aa}, \citet{rukeya_2017_aa}, and \citet{starrfield_2024_aa}.  
For both compositions, we use the abundance distributions tabulated in \citet{kelly_2013_aa} who obtained the ONe mixture from \citet{ritossa_1996_aa}.   

The next two compositions are CO mixtures.  They are 
25\% CO WD matter and 75\% solar (CO 25/75), and 50\% CO WD matter and 50\% solar matter (CO 50/50).  Both compositions were used in \citet{hernanz_1996_aa}, \citet{jose_1998_aa}, \citet{rukeya_2017_aa}, and \citet{starrfield_2020_aa}.  

As in \citet{starrfield_2024_aa}, we use 300 mass zones with the mass of the zone decreasing from the inner boundary to the
surface.  Unlike our earlier studies, however, we assume an initial WD luminosity of $10^{-2}$ \Lsun and choose an \.M  
that results in the WD reaching the TNR in $\approx$80 yr. The initial conditions, mass accretion rate, \.M, the evolutionary time to the TNR, 
and the accreted mass for the solar composition accretion are given in Table \ref{tcrbtable1}.  {The necessary \.M to reduce the evolution time
to achieve a TNR in 80 yr is $\approx 10^{-8}$M$_\odot$yr$^{-1}$ which is about 100 times larger than the values that
we used in our CO and ONe CN studies \citep{starrfield_2020_aa, starrfield_2024_aa} }.

\begin{deluxetable}{@{}lccc}
\tabletypesize{\small}
\tablecaption{Initial Parameters for Solar Accretion onto 1.35 M$_\odot$ White Dwarfs with 3 Different Radii\label{tcrbtable1}}
\tablewidth{0pt}
\tablecolumns{4}
 \tablehead{ \colhead{White Dwarf Radius:} &
 \colhead{1522 km}\tablenotemark{a}&
  \colhead{1827 km}\tablenotemark{b} &
\colhead{2166 km}\tablenotemark{c} }
 \startdata
Inner radius (km)&915&1129&0.0\tablenotemark{d}\\ 
Initial:  L/L$_\odot$($10^{-2}$)&1.0&1.0&1.0\\
Initial: T$_{\rm eff}$($10^4$K)&3.9&3.6&3.3\\
 \.M  ($10^{18}$ gm s$^{-1}$)&0.7&1.0&1.5\\
$\tau$$_{\rm acc}$(yr)\tablenotemark{e}&86.2&80.2&81.4\\
M$_{\rm acc}$($10^{-6}$M$_{\odot}$)\tablenotemark{f}&0.96&1.27&1.93\\
\enddata
\tablenotetext{a}{ \citet[][Table 1]{althaus_2022_ab}, 1.35 M$_\odot$ ONe WD with GR}
\tablenotetext{b}{ \citet[][Table 1]{althaus_2022_ab}, 1.35 M$_\odot$ ONe WD without GR}
\tablenotetext{c}{ \citet[][Table 1]{althaus_2023_aa}, 1.35 M$_\odot$ CO WD without GR}
\tablenotetext{d} { center of the complete WD}
\tablenotetext{e} {Accretion time to the thermonuclear runaway}
\tablenotetext{f}{Accreted mass}
\end{deluxetable}

Also, as in \citet{jose_2007_aa, jose_2020_aa} and \citet{starrfield_2020_aa, starrfield_2024_aa}, we first accrete a solar 
composition \citep{lodders_2003_aa} until a specific rate of energy generation is reached ($\sim 10^{11}$ \erggm ), and then turn off
mass accretion.  We then either continue the simulation with a solar
composition through the peak of the TNR and return to quiescence, or switch to one of the mixed compositions and
follow that simulation through the peak of the TNR and decline. This mimics the behavior seen in multi-D simulations where
WD matter is mixed up into the accreted matter after convection is ongoing and has nearly reached the surface of the
WD \citep[][and references therein]{casanova_2010_aa, casanova_2011_aa, jose_2014_aa}. Detailed descriptions of this procedure as used
in NOVA are found
in \citet{starrfield_2020_aa, starrfield_2024_aa}, where we have described the procedure as Mixing During the
Thermonuclear Runaway (MDTNR).

\section{Results}
\label{results}

 \begin{figure}[htb!]
\includegraphics[width=0.95\textwidth]{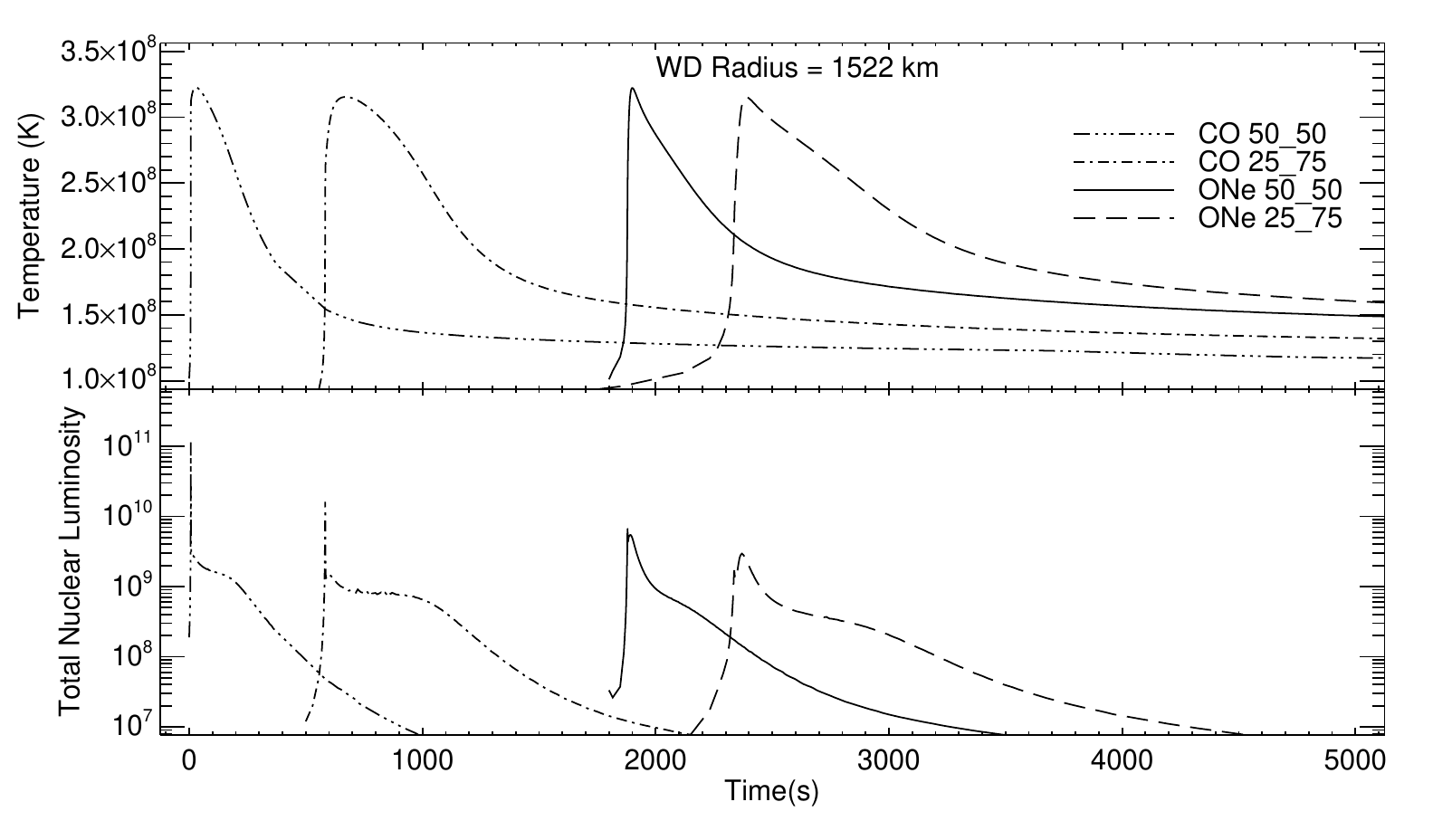}
\caption{Top panel:  the variation with time of the temperature for the mass zones near the interface
between the outer layers of the WD and the accreted matter for the 1522 km WD. The results for four simulations are shown with the WD composition identified in the legend.  The curve for each sequence has been arbitrarily shifted in time to improve its visibility. Interestingly, the peak temperature achieved in each simulation is virtually independent of composition. We do not plot the solar composition simulation because it evolves much more slowly than any of the enriched composition simulations.
Bottom panel: The variation with time of the total nuclear luminosity in solar units (L$_\odot$) around the
time of peak temperature. We integrate over all zones with ongoing nuclear burning to obtain the plotted numbers. The identification with each composition is given in the top panel. The evolution time has been shifted to line up with the temperature variations in the top panel. The irregularities seen in the evolution are caused by the convective region changing its spatial extent, with respect to the mass zones,  as the material expands. In contrast to the results in the top panel, the CO enriched simulations reach higher luminosities than the ONe enriched simulations.}
\label{figure1}
\end{figure}

The evolutionary results for all five compositions are given in Table \ref{tcrbtable2}.  The composition for each of the
simulations is given in the top row.  The next three rows give the initial $^1$H (X), $^4$He (Y), and metals  (Z).
We follow with the particular radius used in the rows below. We begin
with the smallest radius WD, 1522 km. 
The rows tabulate, as a function of WD composition, the peak temperature in the simulation (T$_{\rm peak}$),
the peak rate of energy generation ($\epsilon_{\rm nuc-peak}$), 
the peak surface luminosity in units of the solar luminosity, (L$_{\rm peak}$/L$_\odot$), 
the peak effective temperature (T$_{\rm eff-peak}$), the amount of
mass ejected in solar masses (M$_{\rm ej}$),  and
the amount of \7Li ejected in units of solar mass, $^7$Li$_{\rm ej}$.  Here, we assume that all the $^7$Be produced in the 
TNR decays to \7Li.  The ejected $^7$Be and \7Li abundances (in mass fraction) are given in Tables
\ref{rad1522abund}, \ref{rad1827abund},  and  \ref{rad2166abund} for the three WD radii.

The next two rows give the amount of $^{22}$Na and $^{26}$Al ejected in solar masses.  These two
radioactive elements are predicted to be produced mostly in ONe CNe, but we provide our predictions for 
all 5 compositions.  The final two rows give the Retention Efficiency (RE): (M$_{\rm acc}$ -  M$_{\rm ej}$)/M$_{\rm acc}$,
and the velocity of the surface zone which is the maximum velocity in each simulation (V$_{\rm max}$). 
The RE clearly shows that most of the accreted material remains on the WD surface and is not ejected. Therefore, the
WD is growing in mass as a consequence of the RN phenomenon.  The CO 50/50 simulations, for all 3 radii, eject the most
amount of matter because the increased amount of $^{12}$C produces the highest amount of energy generation and, 
thereby, the most violent outbursts.  The next set of rows shows the same information 
for the WD with a radius of 1827 km, and the final set of rows give the results for the WD radius of 2166 km.

 \begin{deluxetable}{@{}lccccccc}
\tabletypesize{\small}
\tablecaption{Evolutionary results for  Accretion onto a 1.35 M$_\odot$ WD with 3 Different Radii\tablenotemark{a}\label{tcrbtable2}}
\tablewidth{0pt}
\tablecolumns{6}
 \tablehead{ \colhead{Composition:} &
 \colhead{Solar} &
 \colhead{CO 25/75}\tablenotemark{b} &
\colhead{CO 50/50}\tablenotemark{c} &
 \colhead{ONe 25/75}\tablenotemark{d} &
 \colhead{ONe 50/50\tablenotemark{e}}}

 \startdata
 X (initial) &0.71&0.533&0.355&0.533&0.355\\
 Y (initial)&0.274&0.206&0.137&0.206&0.137\\
 Z (initial)&0.015&0.261&0.508&0.261&0.508\\
 \hline
{\bf WD Radius = 1522 km\tablenotemark{f}}&&&&\\
 \hline
T$_{\rm peak}$($10^8$K)&2.8&3.2&3.2&3.2&3.2\\
$\epsilon_{\rm nuc-peak}$(erg gm$^{-1}$s$^{-1}$)&$1.8\times10^{14}$&$1.3 \times10^{17}$&$9.3 \times10^{17}$&$1.5 \times10^{16}$&$6.1\times10^{16}$\\
L$_{\rm peak}$/L$_\odot$ ($10^4$)&4.2&6.3&37.3&4.8&6.2\\
T$_{\rm eff-peak}$($10^6$K)&1.2&1.4&1.5&1.3&1.4\\
M$_{\rm ej}$($10^{-8}$M$_{\odot}$)&$1.3$&$5.6$&$42.0$&$4.6$&5.3\\
$^{7}$Li$_{\rm ej}$(M$_{\odot}$)&$1.4 \times10^{-18}$&$1.2 \times10^{-12}$&$6.3 \times10^{-12}$&$8.1 \times10^{-14}$&$5.5 \times10^{-13}$\\
$^{22}$Na$_{\rm ej}$(M$_{\odot}$)&$2.2 \times10^{-14}$&$1.8 \times10^{-13}$&$1.9 \times10^{-12}$&$3.3 \times10^{-12}$&$1.8 \times10^{-11}$\\
$^{26}$Al$_{\rm ej}$(M$_{\odot}$)&$2.4 \times10^{-15}$&$6.8 \times10^{-14}$&$3.3 \times10^{-12}$&$2.4 \times10^{-12}$&$1.4 \times10^{-11}$\\
RE: (M$_{\rm acc}$ -  M$_{\rm ej}$)/M$_{\rm acc}$&0.99&0.94&0.56&0.95&0.94\\
V$_{\rm max}$(km s$^{-1}$)&536&824&3623&768&880\\
\hline
{\bf WD Radius = 1827 km\tablenotemark{g}}&&&&&&\\
\hline
T$_{\rm peak}$($10^8$K)&2.56&2.80&2.86&2.80&2.86\\
$\epsilon_{\rm nuc-peak}$(erg gm$^{-1}$s$^{-1}$)&$1.5\times10^{14}$&$5.8\times10^{16}$&$3.0 \times10^{17}$&$8.6 \times10^{15}$&$2.5\times10^{17}$\\
L$_{\rm peak}$/L$_\odot$ ($10^4$)&4.1&4.5&46.0&4.6&20.2\\
T$_{\rm eff-peak}$($10^6$K)&1.1&1.3&1.4&1.1&1.2\\
M$_{\rm ej}$($10^{-8}$M$_{\odot}$)&$8.0$&$2.6$&$35.1$&$1.5$&10.7\\
$^{7}$Li$_{\rm ej}$(M$_{\odot}$)&$6.2 \times10^{-18}$&$5.1 \times10^{-13}$&$5.2 \times10^{-12}$&$2.4 \times10^{-14}$&$9.9 \times10^{-13}$\\
$^{22}$Na$_{\rm ej}$(M$_{\odot}$)&$1.9 \times10^{-14}$&$1.1 \times10^{-13}$&$1.1 \times10^{-12}$&$1.7 \times10^{-12}$&$1.4 \times10^{-11}$\\
$^{26}$Al$_{\rm ej}$(M$_{\odot}$)&$1.1 \times10^{-14}$&$1.3 \times10^{-13}$&$3.9\times10^{-12}$&$1.6 \times10^{-12}$&$3.0 \times10^{-11}$\\
RE: (M$_{\rm acc}$ -  M$_{\rm ej}$)/M$_{\rm acc}$&0.94&0.98&0.72&0.99&0.92\\
V$_{\rm max}$(km s$^{-1}$)&639&910&2719&857&1024\\
\hline
{\bf WD Radius = 2166 km\tablenotemark{h}}&&&&&&\\
\hline
T$_{\rm peak}$($10^8$K)&2.49&2.59&2.65&2.59&2.65\\
$\epsilon_{\rm nuc-peak}$(erg gm$^{-1}$s$^{-1}$)&$1.5\times10^{14}$&$3.4 \times10^{16}$&$1.5 \times10^{17}$&$7.0 \times10^{15}$&$2.0\times10^{16}$\\
L$_{\rm peak}$/L$_\odot$ ($10^4$)&4.0&4.6&8.4&5.4&6.3\\
T$_{\rm eff-peak}$($10^6$K)&1.0&1.2&1.2&1.0&1.1\\
M$_{\rm ej}$($10^{-8}$M$_{\odot}$)&$0.0$&$6.0$&$34.9$&$3.2$&1.2\\
$^{7}$Li$_{\rm ej}$(M$_{\odot}$)&$0.0$&$1.2 \times10^{-12}$&$5.0 \times10^{-12}$&$4.0\times10^{-14}$&$1.2 \times10^{-13}$\\
$^{22}$Na$_{\rm ej}$(M$_{\odot}$)&0.0&$1.8 \times10^{-13}$&$8.1 \times10^{-13}$&$1.8 \times10^{-12}$&$4.1 \times10^{-12}$\\
$^{26}$Al$_{\rm ej}$(M$_{\odot}$)&0.0&$5.1 \times10^{-13}$&$1.1 \times10^{-11}$&$3.1 \times10^{-12}$&$4.1 \times10^{-12}$\\
RE: (M$_{\rm acc}$ -  M$_{\rm ej}$)/M$_{\rm acc}$&1.0&0.97&0.82&0.98&0.99\\
V$_{\rm max}$(km s$^{-1}$)&0.0&499&1655&466&578\\
\enddata
\tablenotetext{a}{Radii from \citet{althaus_2022_ab, althaus_2023_aa}}
\tablenotetext{b}{Composition: Carbon-Oxygen Core 25\% and 75\% Solar \citep{lodders_2003_aa}}
\tablenotetext{c}{Composition: Carbon-Oxygen Core 50\% and 50\% Solar \citep{lodders_2003_aa}}
\tablenotetext{d}{Composition: Oxygen-Neon Core 25\% and 75\% \citep{lodders_2003_aa} }
\tablenotetext{e}{Composition: Oxygen-Neon Core 50\% and 50\% Solar \citep{lodders_2003_aa}}
\tablenotetext{f}{\.M = $1.1\times 10^{-8}$ M$_\odot$yr$^{-1}$; $\tau_{\rm{acc}}$=86.2yr; M$_{\rm acc}$ = $9.6\times10^{-7}$M$_{\odot}$}
\tablenotetext{g} {\.M = $1.6\times 10^{-8}$ M$_\odot$yr$^{-1}$; $\tau_{\rm{acc}}$=80.2yr M$_{\rm acc}$ = $1.3\times10^{-6}$M$_{\odot}$}
\tablenotetext{h} {\.M = $2.4\times 10^{-8}$ M$_\odot$yr$^{-1}$; $\tau_{\rm{acc}}$=81.4yr  M$_{\rm acc}$ = $1.9\times10^{-6}$M$_{\odot}$}
\end{deluxetable}

 \begin{figure}[htb!]
\includegraphics[width=1.0\textwidth]{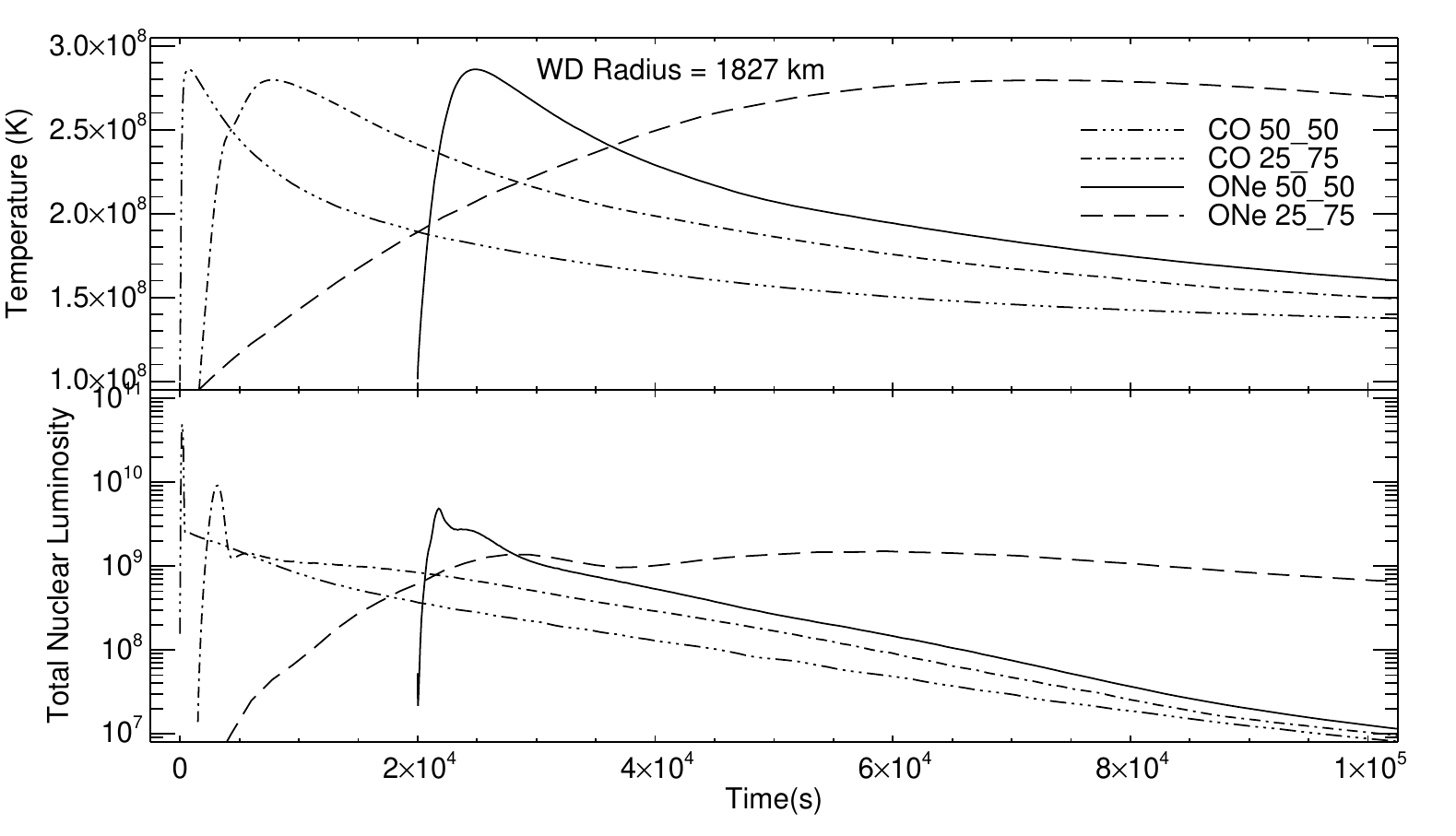}
\caption{Top panel: same as Figure \ref{figure1} (including an arbitrary shift in time to improve visibility) but for the simulations on a WD with a radius of 1827 km. 
Again the peak temperature is nearly independent of composition. In addition, the time scale of the x-axis is
much longer than in Figure \ref{figure1} and the sequences take longer to evolve through the peak of the TNR.
The ONe 25/75 simulation takes about 1 day to reach peak and declines extremely slowly. 
Bottom panel: the same plot as in Figure \ref{figure1} but for the simulations for a WD with a radius of 1827 km. 
Here, the CO 50/50 simulation reaches a higher peak luminosity and evolves much faster than the
other sequences.}
\label{figure2}
\end{figure}

\begin{figure}[htb!]
\includegraphics[width=1.0\textwidth]{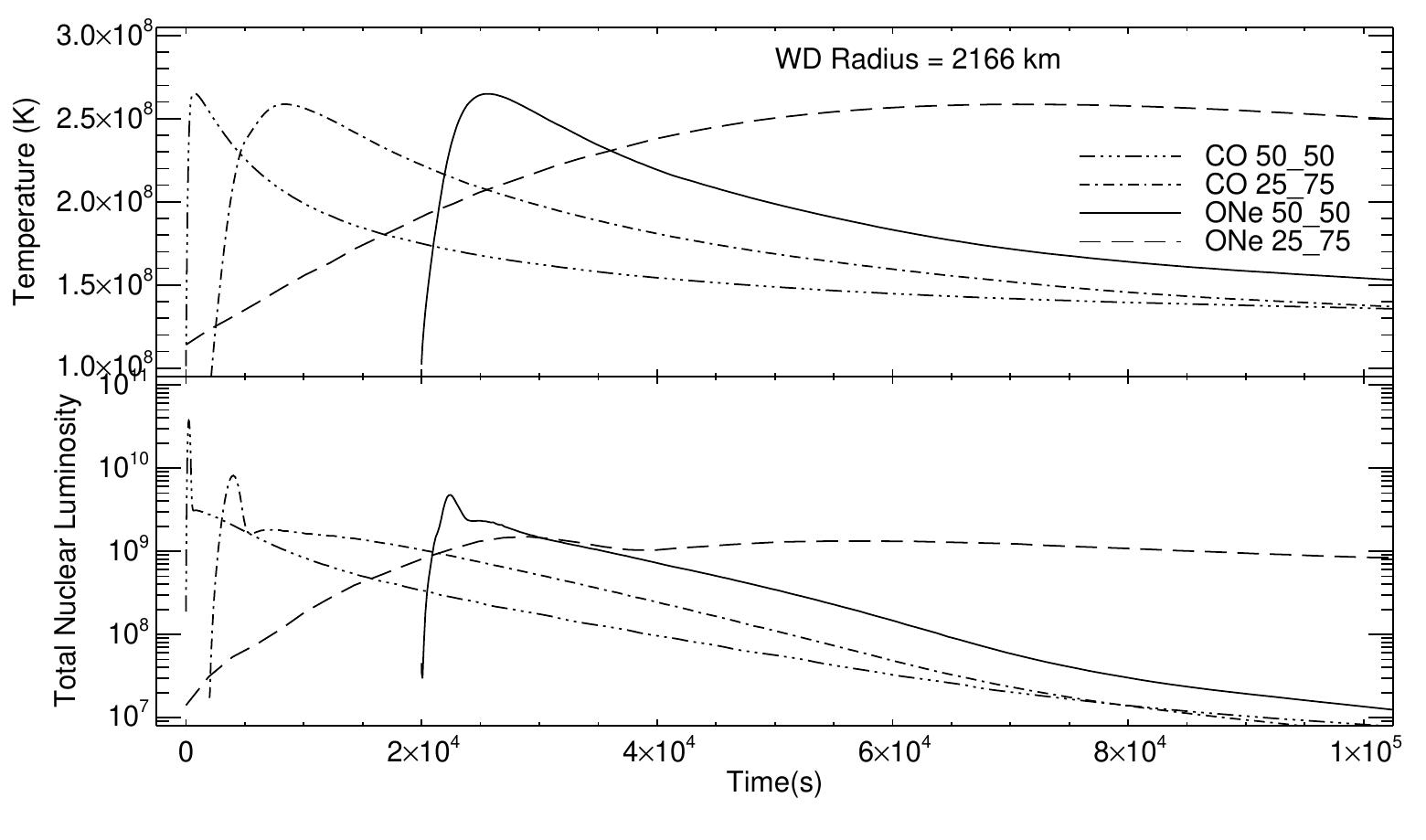}
\caption{Top panel: same as Figure \ref{figure1} (including an arbitrary shift in time to improve visibility)
but for the simulations on a WD with a radius of 2166 km. 
Again the peak temperature is nearly independent of composition. In addition, the time scale of the x-axis is
much longer than in Figure \ref{figure1} and the sequences take longer to evolve through the peak of the TNR.
The ONe 25/75 simulation takes about 1 day to reach peak and declines extremely slowly. 
Bottom panel: the same plot as in Figure \ref{figure1} but for the simulations for a WD with a radius of 2166 km. 
Here, the CO 50/50 simulation reaches a higher peak luminosity and evolves much faster than the
other sequences. }
\label{figure3}
\end{figure}

The plots of temperature versus time for these three sets of simulations are given in the top panels of Figures \ref{figure1}, \ref{figure2}, and \ref{figure3}.  
The simulations with the smallest radius evolve faster (5000 s) than those with the larger radii ($10^5$ s). We do not plot the solar mixture simulations
since they evolve far more slowly then the other simulations and have
not reached maximum temperature by the end of the plot ($10^5$ s). 

\needspace{2 \baselineskip}

\subsection{Evolution of the 1522 km WD}
\label{1522km}

Figure \ref{figure1} and Table \ref{tcrbtable2}  show, for the 1522 km simulations, that the peak temperature
is the same for all the mixed compositions and higher than the peak temperatures for either the simulations of the 1827 km or the 2166 km WDs.
However, peak energy generation depends on the amount of $^{12}$C in the mixture and the highest rate of energy generation occurs for the
CO 50/50 simulation.  

The highest value of T$_{\rm eff}$ occurs just after the positron-decaying nuclei reach the surface as a result of  strong convection, and
the value of T$_{\rm eff-peak}$
is also highest for the CO 50/50 simulation on the 1522 km WD.  
The effects of increasing the abundance of $^{12}$C can also be seen in the bottom panel of Figure \ref{figure1} which shows the evolution of the total nuclear luminosity in solar units (L/L$_\odot$).  The very sharp spikes in both CO simulations are caused
by the proton captures on the enriched $^{12}$C causing it to evolve to both $^{13}$N and $^{14}$O. 
As was realized by \citet{caughlan_1965_aa} and \citet{starrfield_1972_aa}, near peak temperature most of the available proton-capture nuclei become  positron-decaying nuclei, halting any further rise in energy generation.  
All simulations show a rapid rise, in both temperature and nuclear energy generation, to maximum followed by a slower decline.   
The rise to maximum nuclear luminosity occurs as the convective region encompasses all the accreted layers, thus
carrying the  positron-decaying nuclei to the surface and unprocessed nuclei down to the nuclear burning region.

The peak in the luminosity (L$_{\rm peak}$/L$_\odot$) for the 1522 km WD simulation occurs much later in the outburst when the expanding material reaches radii from
$10^{10}$ cm to $10^{11}$ cm or slightly larger. (This is also true for the simulations on the two larger radii WDs.) 
The continued radiation pressure on the outer layers,  as they expand and cool, causes 
them to accelerate to the velocities listed at the bottom of the compilation for each WD radius (Table \ref{tcrbtable2}). The highest velocities are those for the
CO 50/50 simulations for each WD radius, while the lowest velocities are found in the solar simulations.  

The peak luminosities, L$_{\rm peak}$/L$_\odot$, in the simulations for the 1522 km WD are approximately the same except for the CO 50/50 composition.  
Here, the increased amount of $^{12}$C and resulting energy generation have resulted in a more luminous outburst. 
However, peak effective temperature, T$_{\rm eff-peak}$, occurs extremely early in the outburst when the growing convective region reaches the surface, and the positron-decaying nuclei raise the 
energy generation in the surface layers to values exceeding $\sim 10^{13}$ erg gm$^{-1}$s$^{-1}$.  For all radii and all compositions, 
the peak T$_{\rm eff}$ exceeds $10^{6}$ K producing a short phase of X-ray emission.  This prediction of an early ``flash'' of X-ray photons has been found in most of our simulations \citep[][and references therein]{starrfield_1990_aa, starrfield_2024_aa} and has now been confirmed by observations of Nova YZ Reticuli with the eROSITA x-ray satellite \citep{konig_2022_aa}.  

{Concomitant with the X-ray flash is a UV flash that ionizes the red giant wind \citep{shore_1996_aa, shore_1998_aa, munari_2024_aa}.  
The consequences of the UV flash are narrow emission lines superimposed on broad lines in SyRNe such as V407 Cyg, RS Oph, and V3890 Sgr.
The line profiles and a physical description of the evolution, which depends on the 3D nature of the expanding gases, including the recombination time scale, can be found in \citet[][and references therein]{shore_1996_aa, munari_2024_aa}. Since the structure of TCrB is similar to these SyRNe, we expect the
same behavior in the early high dispersion spectra of this RN.  \citet{munari_2024_aa} in his Figure 2, notes that the interaction of the UV flash
and the ionization of the wind increases the light output of the outburst by 2 magnitudes.}

As the expanding layers reach values exceeding $\sim10^{11}$ cm, the continuing acceleration, caused by radiation pressure, drives the outermost layers to speeds greater than the escape speed.  The layers continue to accelerate until they reach radii of $\sim10^{12}$ cm and then begin to decline in velocity as they continue to move out of the WD potential well. 

 We follow the evolution until the densities of the escaping layers drop below the minimum densities in the EOS tables ($10^{-12}$gm cm$^{-3}$).
 Those layers that have become optically thin, and for which their speeds exceed the local escape velocity, are included in the value of ejected mass in solar masses (M$_{\rm ej}$).  The largest amount of ejected mass is seen in the CO 50/50 simulations and that material is moving at the highest velocities. Examining the retention efficiency (RE), we find that in no case is a 
simulation ejecting as much material as was accreted and, therefore, the mass of the WD is growing toward the Chandrasekhar limit.

Since less mass is accreted in the present
simulations, less mass is ejected when compared to the CN modeling in \citet{starrfield_2020_aa, starrfield_2024_aa}.  However, TCrB goes into
outburst far more often than a typical CN and over time will eject as much mass as a CN that outbursts on $\sim10^5$ yr timescales
\citep{gehrz_1998_aa,  darnley_2020_aa, darnley_2021_aa,dellavalle_2020_aa}.

In Section \ref{remnant} we continue the evolution of the WD with the escaping matter removed from the calculations in order to test
the model of \citet{munari_2023_ab} for the cause of the secondary maximum in the TCrB outbursts..

\subsection{Evolution of the 1827 km WD}
\label{1827km}

The next set of rows in Table \ref{tcrbtable2} provides the same information, but for simulations on the 1827 km radius WD.  
Because of the weaker gravitational potential, the simulations for this radius accreted more mass than those on the 1522 km WD (see Table \ref{tcrbtable1}).  
Figure \ref{figure2} shows the variation of the temperature with time for the four mixed compositions at this radius.  T$_{\rm peak}$ 
is only slightly dependent on the composition, and the ONe 25/75 simulation evolves much more slowly than the
other three simulations plotted in this figure.  

The peak temperatures are lower than in the 1522 km WD simulations
because the gravities and densities are lower in the simulations on the larger WDs. 
Because of the larger accreted mass,  the simulation with
the ONe 50/50 composition has a peak energy generation that is four times higher than the same composition on the 1522 km WD. 
The three other mixed composition simulations on this WD show a lower peak energy generation.  The bottom panel in
Figure \ref{figure2} shows the variation with time of the total nuclear luminosity in solar units.  The CO 50/50 mixture evolves much
more rapidly through the peak than the other compositions (note the extremely sharp peak).

The outer layers continue to expand and, in all cases for this radius, some zones eventually reach escape velocity, become optically thin, and are tabulated as ejected.  While the 1827 km WD accreted more mass than the 1522 km WD, only the solar composition and the ONe 50/50 simulations ejected more mass.  The RE shows that, just as for the smaller radius simulations, most of the accreted matter 
remains on the WD and the mass of the WD is increasing.  The final row (Table \ref{tcrbtable2}) gives the velocity of the surface mass zone. The highest velocities for this radius are again for the CO 50/50 mixture, but they are smaller than for the simulation on the 1522 km WD. 
The other four tabulated velocities are larger than the equivalent velocities in the 1522 km WD. This is because the gravity is less at
the larger radius.

\begin{figure}[htb!]
\includegraphics[width=1.0\textwidth]{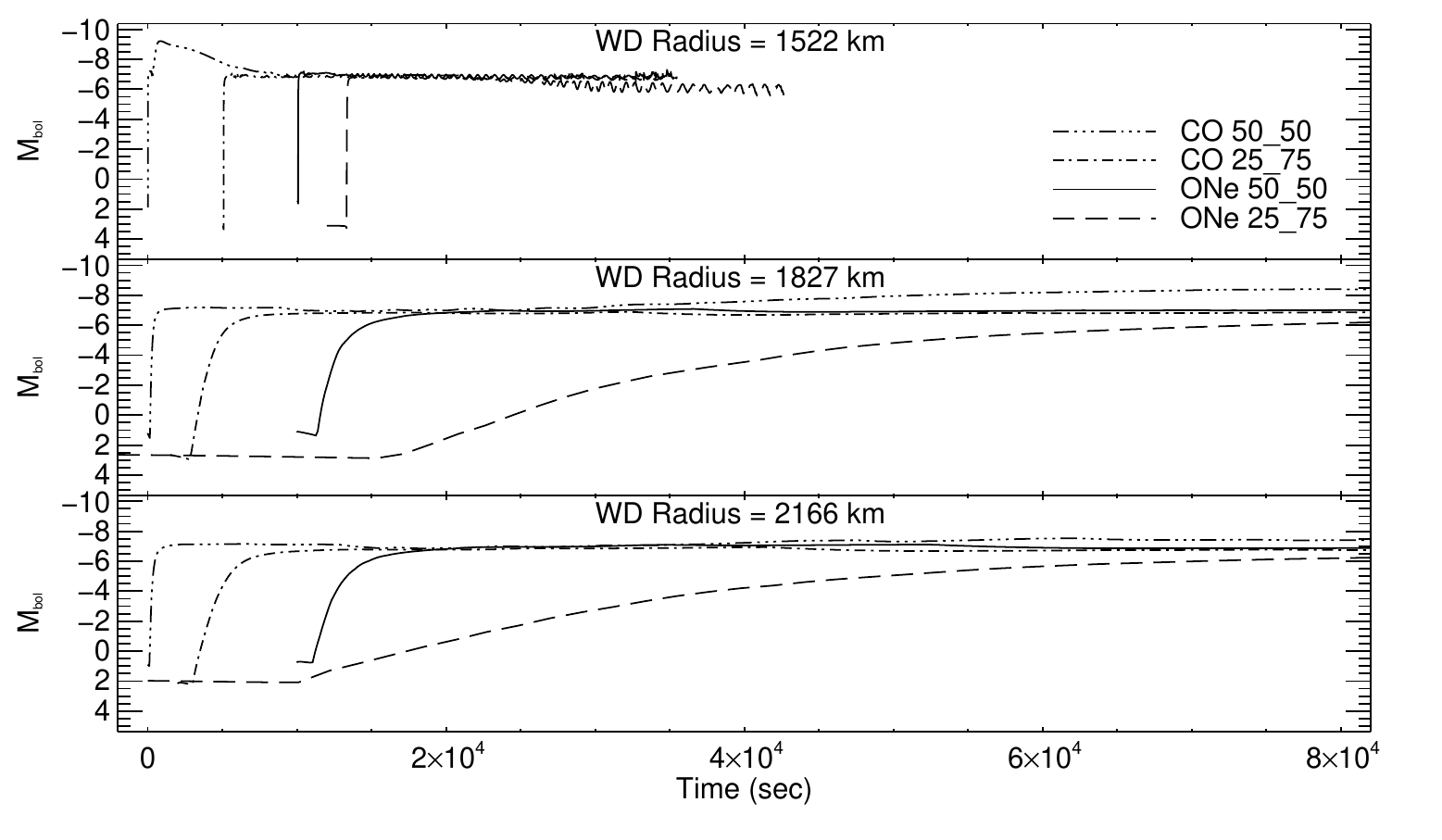}
\caption{The variation with time of the absolute bolometric magnitude for each of the
WDs.  The radius for each set of compositions is given in each panel and the composition is given in the legend.
The top panel shows the light curve evolution for the 1522 km set of simulations which ends early because the
expansion velocities were sufficiently large that the escaping layers reached $\sim 10^{12}$  cm much earlier than the simulations
on the WDs with the larger radii. The oscillations are real.}
\label{figure4}
\end{figure}

\subsection{Evolution of the 2166 km WD}
\label{2166km}

The final set of rows are for the largest radius, the 2166 km CO WD.  In contrast to the two smaller radii, in these simulations we include
the complete WD (i.e., the mass zones extend to the center of the WD).  This radius falls in between the
GR and non-GR radii published by \citet{althaus_2023_aa} for CO WDs. Because the surface gravity is lower, more mass is accreted than for the smaller radii.   However, the density at
the core-envelope interface is smaller and, therefore, the peak temperature is lower than for the simulations on the WDs with smaller radii (c.f., top panel in Figure \ref{figure3}). 
 Peak nuclear energy generation, $\epsilon_{\rm nuc-peak}$, is also lower for these simulations. We show
the total energy generation for all compositions, as a function of time, in the bottom panel of Figure \ref{figure3}. 

The peak effective temperature, T$_{\rm eff-peak}$, is approximately the same for all five compositions and again (as in the smaller radii simulations) reaches values that produce a short episode of X-ray and UV emission.  Comparing the peak  L/L$_\odot$ values for the 2166 km WD simulations to the 1827 km WD simulations, 
we find that the 2166 km WD values are higher for the CO 25/75 and ONe 25/75 simulations, but they are lower 
for the CO 50/50 and ONe 50/50 simulations.  

The solar simulation for this radius ejects no material and, thereby, the RE is 1.0.  As a result, the values of the ejected masses of $^7$Li, $^{22}$Na, and $^{26}$Al are 0.0.  In contrast, the CO 25/75 simulation for  this radius
ejects the most of the three WD radii with the same composition.  Moreover, the CO 50/50 simulations ejects almost as much as the same compositions on the smaller radii
WDs. The ONe 25/75 simulation of this radius ejects more than the same composition for the 1827 km simulation but less
than the 1522 km simulation.  The ONe 50/50 simulation ejects the least amount, of the same composition, on all three WDs. In all cases,
the RE shows that the WDs with a radius of 2166 km are all growing in mass as a consequence of the TCrB outburst.  Finally,
the peak velocities are the lowest of the equivalent simulations on the smaller radii WDs.

\subsection{The Light Curves}
\label{lightcurve}

In Figure \ref{figure4} we plot the light curves for all simulations except the solar simulations, since they evolve too slowly
to plot on the same time scale as the other compositions. The top panel plots M$_{\rm bol}$ versus time for the 1522 km WD. 
These simulations end earlier than those in the lower two panels
because their ejection velocities are higher and the surface layers reach radii exceeding $\sim 10^{12}$ cm earlier than the simulations
on the WDs with larger initial radius (1827 km and 2166 KM).  
 
RNe have been observed to reach bolometric magnitudes of $\sim$-8 \citep{warner_1995_aa, anupama_2008_aa,
darnley_2020_aa, darnley_2021_aa, dellavalle_2020_aa}, but only those for the 1522 km WD simulations
reach or exceed this value.  These are 1-D calculations, however, and thus are unable to simulate the 
observed non-spherical evolution of a real RN.  It is clear from recent reviews that non-spherical
shocks do occur and can increase the observed light from both CNe and RNe \citep[][and references therein]{dellavalle_2020_aa, chomiuk_2021_aa,
munari_2024_aa}.  {In addition, the 3D structure of a SyRN is complex with a great deal of material in the equatorial plane. The ejected gases moving at high speeds will interact with this matter and bypass the smaller amounts of gas in the polar directions.  The consequence of these interactions 
to the evolution and characteristics of the light curves is
discussed in detail in \citet{munari_2024_aa} and not repeated here.}

\begin{figure}[htb!]
\includegraphics[width=1.0\textwidth]{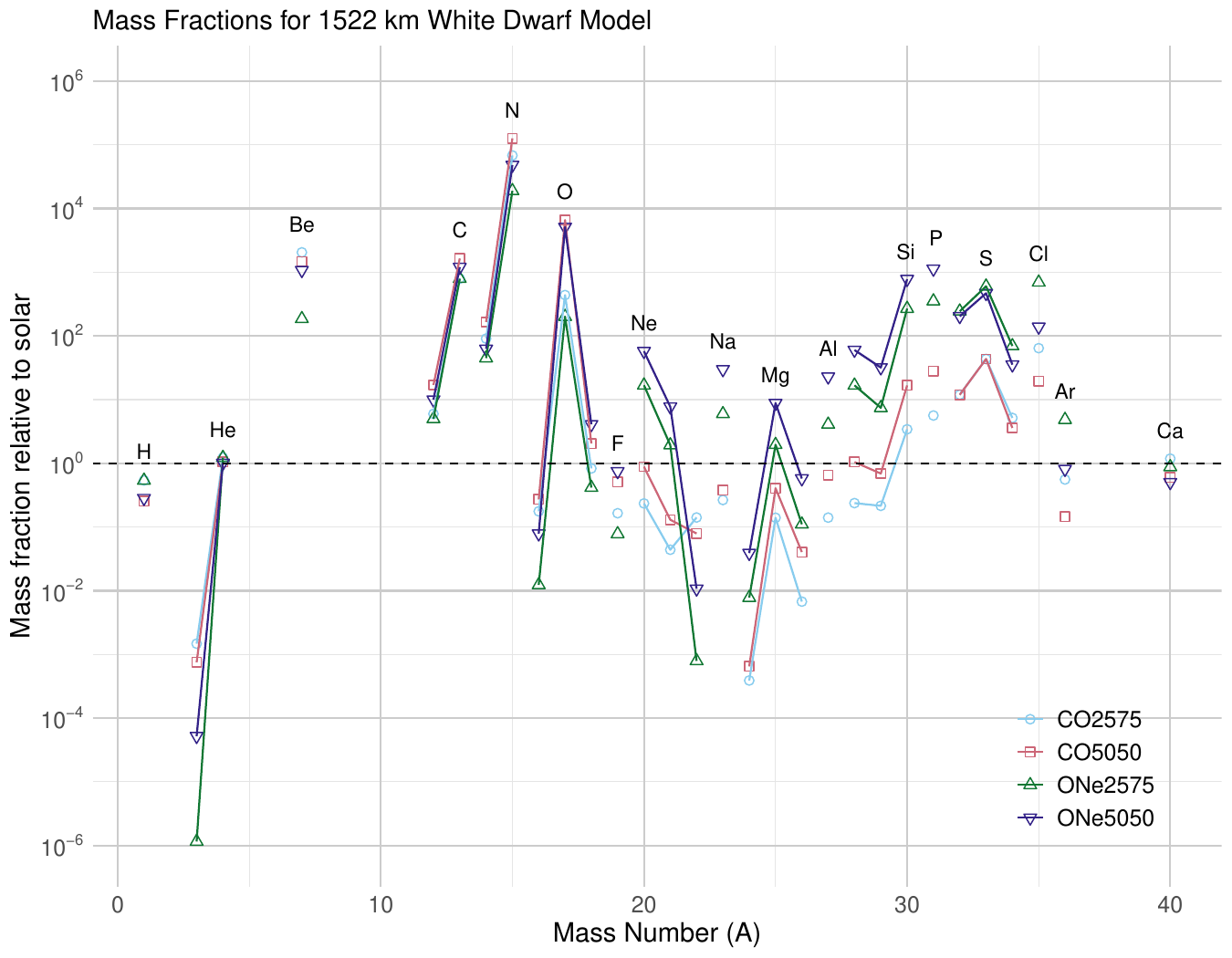}
\caption{The abundances of the stable isotopes from hydrogen to calcium (plus $^7$Be) in the ejecta for the 1522 km WD simulations.  The x-axis is the atomic mass, A, and the y-axis is the logarithmic ratio of the isotopic abundance divided by the solar abundance \citep{lodders_2021_aa}.  $^7$Be is included in this plot because of its large overproduction.  
All isotopes of a given element are connected by solid lines.  We plot the results for all four mixed compositions indicated in the legend.
A number of light, odd isotopes are significantly enriched in the ejecta.  The large depletion of $^3$He is shown and was used, in part, by 
\citet{pepin_2011_aa} to identify ONe nova grains in anomalous interplanetary particles.}
\label{wd1522}
\end{figure}

\needspace{2 \baselineskip}

\section{Nucleosynthesis}
\label{nucleo}

In this section, we emphasize the contributions of a RN such as
TCrB to the abundances of $^7$Li, $^{22}$Na, $^{26}$Al, and $^{31}$P, along with the light, odd isotopes, to
Galactic chemical evolution.  
This section is similar in scope to sections
in \citet{starrfield_2020_aa, starrfield_2024_aa}, for CO and ONe CNe, respectively, where we provided the isotopic results both as
tables of the ejecta abundances and as production plots.  In this paper, Tables \ref{rad1522abund}, \ref{rad1827abund},  and  
\ref{rad2166abund} allow us to compare the ejecta abundance 
predictions for the simulations discussed in the last section (Section \ref{results}) as a function of WD radius. 
The columns list the isotopes down the left hand column and the isotopic predictions (in mass fraction) for each composition are given in the other columns.  The chosen composition is given at the top of each column and the WD radius is given in the table caption.  	
We give the abundance predictions {\it in solar masses}, for $^7$Li, $^{22}$Na, and $^{26}$Al, for each WD radius, in Table \ref{tcrbtable2}.

\begin{deluxetable}{@{}lcccccc}
\tablecaption{Isotopic Ejecta Abundances for 1.35M$_\odot$ WD with a Radius of 1522 km \tablenotemark{a} \label{rad1522abund}}
\tablewidth{0pt}
\tablecolumns{6}
 \tablehead{ \colhead{Composition:} &
 \colhead{Solar} &
 \colhead{CO 25/75}\tablenotemark{b} &
\colhead{CO 50/50}\tablenotemark{c} &
 \colhead{ONe 25/75}\tablenotemark{d} &
 \colhead{ONe 50/50\tablenotemark{e}}}
\startdata
$^1$H\tablenotemark{f}&$6.1 \times10^{-1}$&$3.8 \times10^{-1}$&$1.8 \times10^{-1}$&$3.8 \times10^{-1}$&$2.0 \times10^{-1}$ \\
$^{3}$He&$1.2 \times10^{-8}$  &$5.1 \times 10^{-8}$&$2.6 \times 10^{-8}$& $4.0 \times 10^{-11}$&$1.8 \times 10^{-9}$\\
$^4$He&$3.7 \times10^{-1}$&$3.4 \times10^{-1}$&$2.9 \times10^{-1}$&$3.4 \times10^{-1}$&$2.8\times10^{-1}$\\
$^{7}$Be&$1.4 \times10^{-10}$  &$2.1 \times 10^{-5}$&$1.5 \times 10^{-5}$& $1.9 \times 10^{-6}$&$1.1\times 10^{-5}$\\
$^{7}$Li  &$6.8 \times10^{-14}$&$3.9 \times 10^{-13}$&$1.2 \times 10^{-13}$& $5.7\times 10^{-14}$&$3.0 \times 10^{-13}$\\
$^{12}$C &$1.0 \times10^{-3}$&$1.8 \times10^{-2}$&$5.1 \times 10^{-2}$& $1.5 \times 10^{-2}$&$3.0 \times 10^{-2}$\\
$^{13}$C &$1.9 \times10^{-3}$&$2.7 \times10^{-2}$&$5.6 \times10^{-2}$ &$2.7 \times 10^{-2}$&$4.1 \times 10^{-2}$\\
$^{14}$N&$3.4 \times10^{-3}$ &$7.7 \times 10^{-2}$&$1.4 \times10^{-1}$&$3.8\times10^{-2}$&$5.3 \times10^{-2}$\\
$^{15}$N &$3.0 \times10^{-3}$ &$1.4 \times10^{-1}$&$2.6 \times10^{-1}$& $3.9 \times 10^{-2}$&$9.9 \times10^{-2}$ \\
$^{16}$O &$2.6 \times10^{-5}$ &$1.3 \times10^{-3}$&$2.0 \times10^{-3}$&$9.1 \times10^{-5}$ &$5.8 \times10^{-4}$\\
$^{17}$O &$9.3\times10^{-6}$ &$1.2 \times 10^{-3}$&$1.8 \times 10^{-2}$& $5.5 \times 10^{-4}$&$1.4 \times 10^{-2}$\\
$^{18}$O &$2.9 \times10^{-8}$ &$1.3 \times 10^{-5}$&$3.2 \times 10^{-5}$& $6.5 \times 10^{-6}$&$6.4 \times 10^{-5}$ \\
$^{18}$F &$2.6 \times10^{-9}$ &$8.4 \times 10^{-7}$&$1.8 \times 10^{-5}$& $3.7 \times 10^{-7}$&$4.7 \times 10^{-6}$\\
$^{19}$F &$2.7 \times10^{-10}$ &$1.1 \times 10^{-7}$&$3.4 \times 10^{-7}$& $5.2 \times 10^{-8}$&$5.0 \times 10^{-7}$ \\
$^{20}$Ne &$8.1 \times10^{-5}$ &$5.3 \times 10^{-4}$&$2.0 \times 10^{-3}$& $3.8 \times 10^{-2}$&$1.3 \times 10^{-1}$ \\
$^{21}$Ne &$4.0 \times10^{-8}$ &$2.5 \times 10^{-7}$&$7.4 \times 10^{-7}$& $1.1 \times 10^{-5}$&$4.4 \times 10^{-5}$\\
$^{22}$Ne&$5.9 \times10^{-7}$ &$2.5 \times 10^{-5}$&$1.4 \times 10^{-5}$& $1.4 \times 10^{-7}$&$1.9 \times 10^{-6}$\\
$^{22}$Na &$1.7 \times10^{-6}$ &$3.2 \times 10^{-6}$&$4.6 \times 10^{-6}$& $7.2 \times 10^{-5}$&$3.4 \times 10^{-4}$\\
$^{23}$Na&$5.2 \times10^{-6}$ &$9.8 \times 10^{-6}$&$1.4 \times 10^{-5}$& $2.2 \times 10^{-4}$&$1.1 \times 10^{-3}$ \\
$^{24}$Mg &$2.2 \times10^{-7}$ &$2.1 \times 10^{-7}$&$3.5 \times 10^{-7}$& $4.2 \times 10^{-6}$&$2.1 \times 10^{-5}$ \\
$^{25}$Mg &$2.6 \times10^{-6}$ &$1.0 \times 10^{-5}$&$2.9 \times 10^{-5}$& $1.4 \times 10^{-4}$&$6.3 \times 10^{-4}$\\
$^{26}$Mg&$1.6 \times10^{-7}$ &$5.5 \times 10^{-7}$&$3.3 \times 10^{-6}$& $9.0 \times 10^{-6}$&$4.7 \times 10^{-5}$\\
$^{26}$Al &$2.0 \times10^{-7}$&$1.2 \times 10^{-6}$&$7.8 \times 10^{-6}$& $5.2 \times 10^{-5}$&$2.6 \times 10^{-4}$\\
$^{27}$Al &$1.6 \times10^{-6}$&$8.6 \times 10^{-6}$&$4.0 \times 10^{-5}$& $2.5 \times 10^{-4}$&$1.4 \times 10^{-3}$\\
$^{28}$Si &$7.5 \times10^{-5}$&$1.7 \times 10^{-4}$&$7.5 \times 10^{-4}$& $1.2 \times 10^{-2}$&$4.3 \times 10^{-2}$\\
$^{29}$Si &$4.1 \times10^{-6}$&$8.1 \times 10^{-6}$&$2.6 \times 10^{-5}$& $2.8 \times 10^{-4}$&$1.2 \times 10^{-3}$\\
$^{30}$Si &$3.9 \times10^{-5}$&$8.8 \times 10^{-5}$&$4.3 \times 10^{-4}$& $6.9 \times 10^{-3}$&$2.0 \times 10^{-2}$\\
$^{31}$P &$1.1 \times10^{-5}$&$4.0 \times 10^{-5}$&$2.0 \times 10^{-4}$& $2.5 \times 10^{-3}$&$8.0 \times 10^{-3}$\\
$^{32}$S &$3.2 \times10^{-3}$&$4.4 \times 10^{-3}$&$4.4 \times 10^{-3}$& $9.0 \times 10^{-2}$&$7.5 \times 10^{-2}$\\
$^{33}$S &$6.4 \times10^{-6}$&$1.3 \times 10^{-4}$&$1.3 \times 10^{-4}$& $1.8\times 10^{-3}$&$1.4 \times 10^{-3}$\\
$^{34}$S &$5.6\times10^{-6}$&$8.9 \times 10^{-5}$&$6.2 \times 10^{-5}$& $1.2 \times 10^{-3}$&$6.1 \times 10^{-4}$\\
$^{35}$Cl &$4.1 \times10^{-5}$&$2.5 \times 10^{-4}$&$7.6 \times 10^{-5}$& $2.7 \times 10^{-3}$&$5.4 \times 10^{-4}$\\
$^{36}$Ar &$4.6 \times10^{-6}$&$4.6 \times 10^{-5}$&$1.2 \times 10^{-5}$& $4.0 \times 10^{-4}$&$6.7 \times 10^{-5}$\\
$^{40}$Ca &$1.1 \times10^{-3}$&$7.3 \times 10^{-5}$&$3.7 \times 10^{-5}$& $5.4 \times 10^{-5}$&$3.1 \times 10^{-5}$\\
\enddata
\tablenotetext{a}{Radii from \citet{althaus_2022_ab, althaus_2023_aa}}
\tablenotetext{b}{ Composition: Carbon-Oxygen Core 25\% and 75\% Solar}
\tablenotetext{c}{ Composition: Carbon-Oxygen Core 50\% and 50\% Solar}
\tablenotetext{d}{ Composition: Oxygen-Neon Core 25\% and 75\% Solar}
\tablenotetext{e}{ Composition: Oxygen-Neon Core 50\% and 50\% Solar.}
\tablenotetext{f}{All abundances are mass fraction}
\end{deluxetable}

\begin{figure}[htb!]
\includegraphics[width=1.0\textwidth]{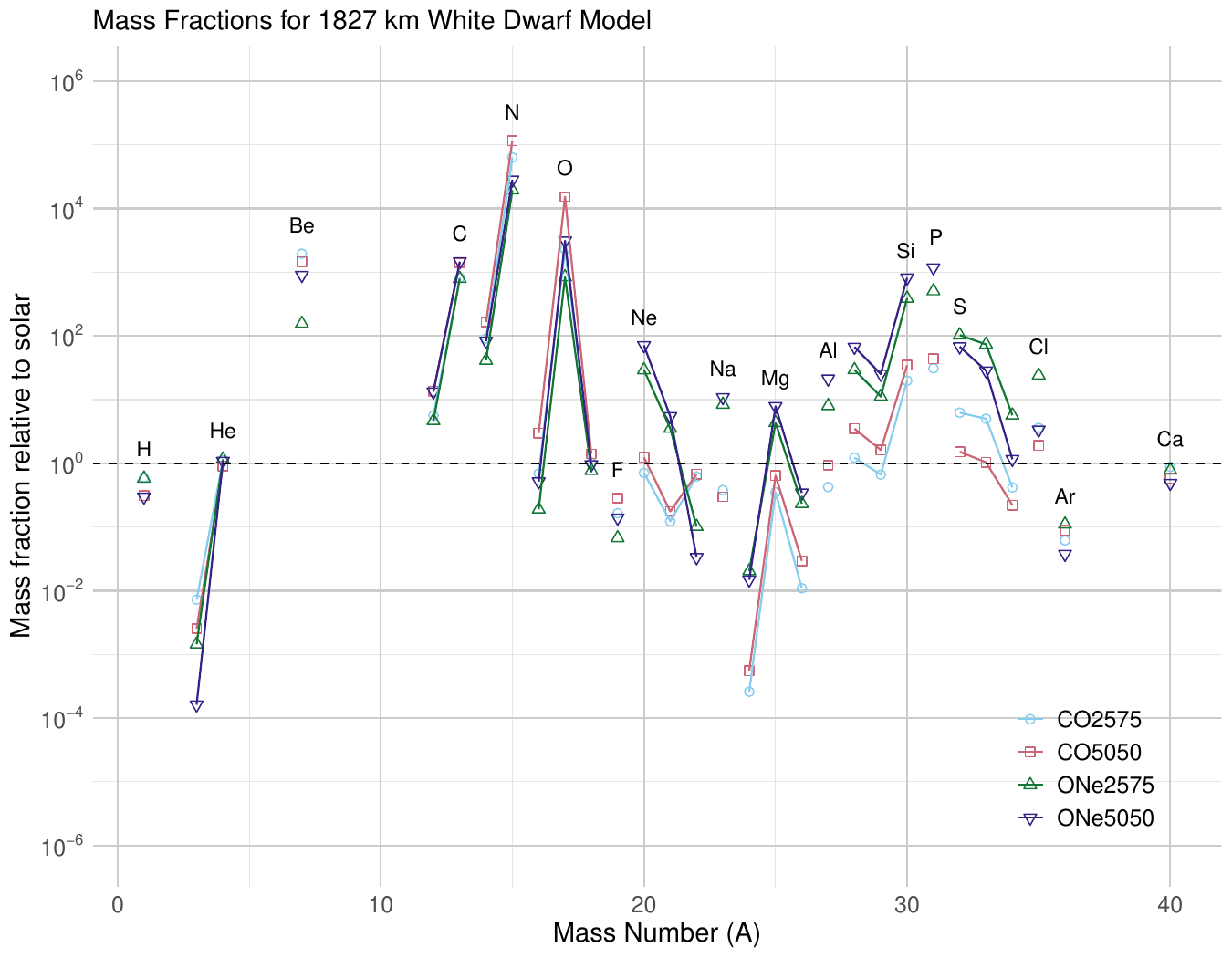}
\caption{Same plot as Figure \ref{wd1522} but for the 1827 km WD. Again, the same isotopes are plotted.  The most striking results are the enrichment of $^7$Be, the enrichment of the odd isotopes: $^{13}$C, $^{15}$N, and $^{17}$O, and the enrichments of $^{30}$ Si and $^{31}$P.}
\label{wd1827}
\end{figure}

\begin{figure}[htb!]
\includegraphics[width=1.0\textwidth]{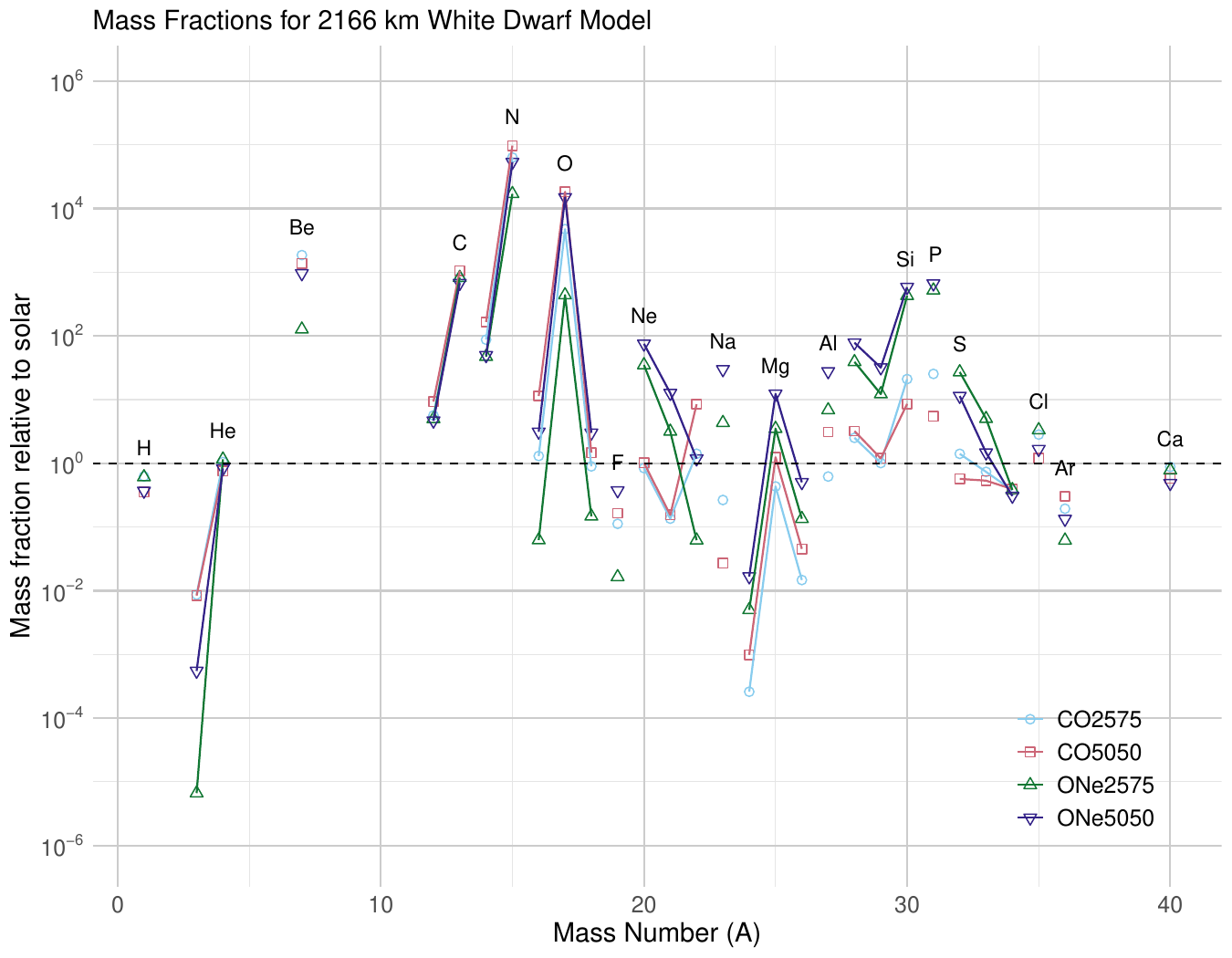}
\caption{ Same plot as Figure \ref{wd1522} but for the 2166 km WD simulations.  Again, the most striking results are the enrichment of $^7$Be (which
does not depend on WD radius), the enrichment of the odd isotopes: $^{13}$C, $^{15}$N, and $^{17}$O, and the enrichment of $^{30}$Si and $^{31}$P.  }
\label{wd2166}
\end{figure}

Figures \ref{wd1522}, \ref{wd1827}, and \ref{wd2166}
provide the abundances of the stable isotopes from hydrogen to calcium (plus $^7$Be with a half-life of  $\sim$ 53 days so it decays
to $^7$Li after we end the simulations) in the ejecta. 
The x-axis is the atomic mass number, A, and the y-axis is the logarithmic ratio of the isotopic abundance divided by the solar abundance \citep{lodders_2021_aa}.  While the solar composition used in the simulations is from \citet{lodders_2003_aa} for ease in
comparison with previous work, it is appropriate to use the modern values from \citet{lodders_2021_aa} for these comparisons.
A detailed description of the methods for obtaining the solar abundances and comparison with previous compilations is
given in \citet{lodders_2021_aa}.
 All isotopes of a given element are connected by solid lines.  The particular element is listed and each of the 4 mixed
compositions is indicated by a different color as shown in the figure legend.
Any isotope above 1.0 is overproduced, and a number of isotopes are significantly enriched in the ejecta. 
Figure \ref{wd1522} provides the comparisons for the mixed compositions on the 1522 km WD, Figure \ref{wd1827} shows the ejecta
isotopes for the 1827 km WD, and Figure \ref{wd2166} shows the same values for the 2166 km WD. 

\begin{deluxetable}{@{}lccccc}
\tablecaption{Isotopic Ejecta Abundances for 1.35M$_\odot$ WD with a Radius of 1827 km \tablenotemark{a} 
\label{rad1827abund}}
\tablewidth{0pt}
\tablecolumns{6}
 \tablehead{ \colhead{Composition:} &
 \colhead{Solar} &
 \colhead{CO 25/75}\tablenotemark{b} &
\colhead{CO 50/50}\tablenotemark{c} &
 \colhead{ONe 25/75}\tablenotemark{d} &
 \colhead{ONe 50/50\tablenotemark{e}}}
\startdata
$^1$H\tablenotemark{f}&$5.9 \times10^{-1}$&$4.2 \times10^{-1}$&$2.2 \times10^{-1}$&$4.1 \times10^{-1}$&$2.1 \times10^{-1}$ \\
$^{3}$He&$1.2 \times10^{-8}$ &$2.5 \times 10^{-7}$&$8.8 \times 10^{-8}$& $5.0 \times 10^{-8}$&$5.6 \times 10^{-9}$\\
$^4$He&$3.9 \times10^{-1}$&$3.1\times10^{-1}$&$2.5 \times10^{-1}$&$3.2 \times10^{-1}$&$3.0\times10^{-1}$\\
$^{7}$Be&$7.8 \times10^{-11}$  &$2.0 \times 10^{-5}$&$1.5 \times 10^{-5}$& $1.6 \times 10^{-6}$&$9.2\times 10^{-6}$\\
$^{7}$Li  &$1.9 \times10^{-12}$&$4.9 \times 10^{-13}$&$1.5 \times 10^{-13}$& $3.9\times 10^{-12}$&$6.3 \times 10^{-13}$\\
$^{12}$C &$1.4 \times10^{-3}$&$1.7 \times10^{-2}$&$4.1 \times 10^{-2}$& $1.4 \times 10^{-2}$&$4.0 \times 10^{-2}$\\
$^{13}$C &$2.4 \times10^{-3}$&$2.8 \times10^{-2}$&$4.8 \times10^{-2}$ &$2.7 \times 10^{-2}$&$5.0 \times 10^{-2}$\\
$^{14}$N&$3.7\times10^{-3}$ &$7.7 \times 10^{-2}$&$1.4 \times10^{-1}$&$3.5\times10^{-2}$&$7.0 \times10^{-2}$\\
$^{15}$N &$1.6 \times10^{-3}$ &$1.3 \times10^{-1}$&$2.4 \times10^{-1}$& $4.0 \times 10^{-2}$&$5.8 \times10^{-2}$ \\
$^{16}$O &$3.7 \times10^{-5}$ &$5.1\times10^{-3}$&$2.2 \times10^{-2}$&$1.4 \times10^{-3}$ &$3.8 \times10^{-3}$\\
$^{17}$O &$4.0\times10^{-6}$ &$6.3 \times 10^{-3}$&$4.2 \times 10^{-2}$& $2.3 \times 10^{-3}$&$8.5 \times 10^{-3}$\\
$^{18}$O &$1.0 \times10^{-8}$ &$1.6 \times 10^{-5}$&$2.2 \times 10^{-5}$& $1.2 \times 10^{-5}$&$1.5 \times 10^{-5}$ \\
$^{18}$F &$4.6 \times10^{-10}$ &$1.8 \times 10^{-6}$&$8.3 \times 10^{-6}$& $9.1 \times 10^{-7}$&$7.2 \times 10^{-7}$\\
$^{19}$F &$1.9 \times10^{-10}$ &$1.1 \times 10^{-7}$&$1.9 \times 10^{-7}$& $4.5 \times 10^{-8}$&$9.3 \times 10^{-8}$ \\
$^{20}$Ne &$6.9\times10^{-5}$ &$1.6\times 10^{-3}$&$2.8 \times 10^{-3}$& $6.6 \times 10^{-2}$&$1.6 \times 10^{-1}$ \\
$^{21}$Ne &$2.0 \times10^{-8}$ &$7.0\times 10^{-7}$&$1.0 \times 10^{-6}$& $2.0 \times 10^{-5}$&$3.1 \times 10^{-5}$\\
$^{22}$Ne&$3.0 \times10^{-7}$ &$1.1 \times 10^{-4}$&$1.2 \times 10^{-4}$& $1.8 \times 10^{-5}$&$5.9 \times 10^{-6}$\\
$^{22}$Na &$2.4 \times10^{-7}$ &$4.4 \times 10^{-6}$&$3.2 \times 10^{-6}$& $1.1 \times 10^{-4}$&$1.3 \times 10^{-4}$\\
$^{23}$Na&$7.3 \times10^{-7}$ &$1.4 \times 10^{-5}$&$1.1 \times 10^{-5}$& $3.1 \times 10^{-4}$&$4.0 \times 10^{-4}$ \\
$^{24}$Mg &$2.1 \times10^{-7}$ &$1.4 \times 10^{-7}$&$3.0 \times 10^{-7}$& $1.1 \times 10^{-5}$&$8.0 \times 10^{-6}$ \\
$^{25}$Mg &$1.1 \times10^{-6}$ &$2.5 \times 10^{-5}$&$4.6 \times 10^{-5}$& $3.1 \times 10^{-4}$&$5.6 \times 10^{-4}$\\
$^{26}$Mg&$1.0 \times10^{-7}$ &$8.9 \times 10^{-7}$&$2.4 \times 10^{-6}$& $1.9 \times 10^{-5}$&$2.8 \times 10^{-5}$\\
$^{26}$Al &$1.3 \times10^{-7}$&$4.9 \times 10^{-6}$&$1.1 \times 10^{-5}$& $1.1 \times 10^{-4}$&$2.8 \times 10^{-4}$\\
$^{27}$Al &$7.2 \times10^{-7}$&$2.6 \times 10^{-5}$&$5.7 \times 10^{-5}$& $4.9 \times 10^{-4}$&$1.3 \times 10^{-3}$\\
$^{28}$Si &$5.2 \times10^{-5}$&$8.8 \times 10^{-4}$&$2.5 \times 10^{-3}$& $2.1 \times 10^{-2}$&$4.8 \times 10^{-2}$\\
$^{29}$Si &$1.0 \times10^{-6}$&$2.5 \times 10^{-5}$&$6.1 \times 10^{-5}$& $4.2 \times 10^{-4}$&$9.5 \times 10^{-4}$\\
$^{30}$Si &$4.8 \times10^{-5}$&$5.1\times 10^{-4}$&$8.9 \times 10^{-4}$& $1.0 \times 10^{-2}$&$2.1 \times 10^{-2}$\\
$^{31}$P &$5.0 \times10^{-6}$&$2.2 \times 10^{-4}$&$3.1 \times 10^{-4}$& $3.6 \times 10^{-3}$&$8.4 \times 10^{-3}$\\
$^{32}$S &$3.4 \times10^{-3}$&$2.3 \times 10^{-3}$&$5.6 \times 10^{-4}$& $3.8 \times 10^{-2}$&$2.5 \times 10^{-2}$\\
$^{33}$S &$6.3 \times10^{-6}$&$1.5 \times 10^{-5}$&$3.1 \times 10^{-6}$& $2.2 \times 10^{-4}$&$8.5 \times 10^{-5}$\\
$^{34}$S &$5.8\times10^{-6}$&$7.2 \times 10^{-6}$&$3.8 \times 10^{-6}$& $9.8 \times 10^{-5}$&$2.0 \times 10^{-5}$\\
$^{35}$Cl &$4.3 \times10^{-5}$&$1.4 \times 10^{-5}$&$7.4 \times 10^{-6}$& $9.4 \times 10^{-5}$&$1.3 \times 10^{-5}$\\
$^{36}$Ar &$4.0 \times10^{-6}$&$5.1 \times 10^{-6}$&$7.3 \times 10^{-6}$& $9.1 \times 10^{-6}$&$3.1 \times 10^{-6}$\\
$^{40}$Ca &$8.1 \times10^{-4}$&$5.4 \times 10^{-5}$&$3.6 \times 10^{-5}$& $4.8 \times 10^{-5}$&$3.0 \times 10^{-5}$\\
\enddata
\tablenotetext{a}{Radii from \citet{althaus_2022_ab, althaus_2023_aa}}
\tablenotetext{b}{ Composition: Carbon-Oxygen Core 25\% and 75\% Solar}
\tablenotetext{c}{ Composition: Carbon-Oxygen Core 50\% and 50\% Solar}
\tablenotetext{d}{ Composition: Oxygen-Neon Core 25\% and 75\% Solar}
\tablenotetext{e}{ Composition: Oxygen-Neon Core 50\% and 50\% Solar.}
\tablenotetext{f}{All abundances are mass fraction}
\end{deluxetable}

Examining the isotopic predictions for the 1522 km WD given in Table \ref{rad1522abund} and Figure \ref{wd1522}, we find
that the major differences are in the intermediate mass nuclei which are all enriched in the ONe 25/75 simulation when compared to the CO 25/75 simulation.  
As already reported in the ONe CN studies \citep{starrfield_2009_aa, starrfield_2024_aa},  
$^3$He is significantly depleted in the ONe simulations.  
$^{13}$C, $^{15}$N,  and $^{17}$O exhibit approximately the same enrichment as $^7$Be but $^{31}$P is more enriched in
the ONe 25/75 simulation than in the CO 25/75 simulation.   Again, as seen in the ONe simulations reported in \citet{starrfield_2024_aa}, the intermediate mass elements are more enriched in the
ONe 25/75 and ONe 50/50 simulations than in the CO simulations.  However, $^{24}$Mg and $^{26}$Mg are significantly
depleted in all four simulations while $^{25}$Mg is enriched in the ONe 50/50 simulation.

\begin{deluxetable}{@{}lccccc}
\tablecaption{Isotopic Ejecta Abundances for 1.35M$_\odot$ WD with a Radius of 2166 km \tablenotemark{a} 
\label{rad2166abund}}
\tablewidth{0pt}
\tablecolumns{6}
 \tablehead{ \colhead{Composition:} &
 \colhead{Solar\tablenotemark{b}} &
 \colhead{CO 25/75}\tablenotemark{c} &
\colhead{CO 50/50}\tablenotemark{d} &
 \colhead{ONe 25/75}\tablenotemark{e} &
 \colhead{ONe 50/50\tablenotemark{f}}}
\startdata
$^1$H\tablenotemark{g}&$6.4\times10^{-1}$&$4.4 \times10^{-1}$&$2.5 \times10^{-1}$&$4.3 \times10^{-1}$&$2.6 \times10^{-1}$ \\
$^{3}$He&$7.2 \times10^{-13}$ &$3.0 \times 10^{-7}$&$2.9 \times 10^{-7}$& $2.3 \times 10^{-10}$&$1.9 \times 10^{-8}$\\
$^4$He&$3.4 \times10^{-1}$&$2.9\times10^{-1}$&$2.1 \times10^{-1}$&$3.2 \times10^{-1}$&$2.3 \times10^{-1}$\\
$^{7}$Be&$1.5 \times10^{-11}$  &$1.9 \times 10^{-5}$&$1.4 \times 10^{-5}$& $1.3 \times 10^{-6}$&$9.8 \times 10^{-6}$\\
$^{7}$Li  &$0.0\tablenotemark{h}$&$6.4 \times 10^{-13}$&$1.9 \times 10^{-13}$& $6.0\times 10^{-14}$&$3.0 \times 10^{-13}$\\
$^{12}$C &$7.5 \times10^{-4}$&$1.7 \times10^{-2}$&$2.8 \times 10^{-2}$& $1.5 \times 10^{-2}$&$1.4 \times 10^{-2}$\\
$^{13}$C &$8.9 \times10^{-4}$&$2.8 \times10^{-2}$&$3.6 \times10^{-2}$ &$2.8 \times 10^{-2}$&$2.3 \times 10^{-2}$\\
$^{14}$N&$3.4 \times10^{-3}$ &$7.4 \times 10^{-2}$&$1.4 \times10^{-1}$&$4.0 \times10^{-2}$&$4.2 \times10^{-2}$\\
$^{15}$N &$4.6 \times10^{-3}$ &$1.3 \times10^{-1}$&$2.0 \times10^{-1}$& $3.5 \times 10^{-2}$&$1.1 \times10^{-1}$ \\
$^{16}$O &$6.5 \times10^{-6}$ &$9.6 \times10^{-3}$&$8.4 \times10^{-2}$&$4.6 \times10^{-4}$ &$2.3 \times10^{-2}$\\
$^{17}$O &$1.2 \times10^{-6}$ &$1.3 \times 10^{-2}$&$5.0 \times 10^{-2}$& $1.2  \times 10^{-3}$&$4.1 \times 10^{-2}$\\
$^{18}$O &$8.8 \times10^{-10}$ &$1.4 \times 10^{-5}$&$2.3 \times 10^{-5}$& $2.3 \times 10^{-6}$&$4.7 \times 10^{-5}$ \\
$^{18}$F &$5.2 \times10^{-11}$ &$8.8 \times 10^{-7}$&$1.7 \times 10^{-6}$& $1.3 \times 10^{-7}$&$3.0 \times 10^{-6}$\\
$^{19}$F &$2.0 \times10^{-11}$ &$7.5 \times 10^{-8}$&$1.1 \times 10^{-7}$& $1.1 \times 10^{-8}$&$2.5 \times 10^{-7}$ \\
$^{20}$Ne &$1.4 \times10^{-4}$ &$1.9 \times 10^{-3}$&$2.3 \times 10^{-3}$& $7.9 \times 10^{-2}$&$1.7 \times 10^{-1}$ \\
$^{21}$Ne &$5.6 \times10^{-8}$ &$7.7 \times 10^{-7}$&$8.8 \times 10^{-7}$& $1.8 \times 10^{-5}$&$7.2 \times 10^{-5}$\\
$^{22}$Ne&$2.2 \times10^{-9}$ &$2.5 \times 10^{-4}$&$1.5 \times 10^{-3}$& $1.1 \times 10^{-5}$&$2.1 \times 10^{-4}$\\
$^{22}$Na &$4.8 \times10^{-7}$ &$3.0 \times 10^{-6}$&$2.3 \times 10^{-6}$& $5.7 \times 10^{-5}$&$3.4 \times 10^{-4}$\\
$^{23}$Na&$1.6 \times10^{-6}$ &$9.8 \times 10^{-6}$&$1.0 \times 10^{-6}$& $1.6 \times 10^{-4}$&$1.1 \times 10^{-3}$ \\
$^{24}$Mg &$1.2 \times10^{-8}$ &$1.4 \times 10^{-7}$&$5.3 \times 10^{-7}$& $2.7 \times 10^{-6}$&$9.0 \times 10^{-6}$ \\
$^{25}$Mg &$8.6 \times10^{-7}$ &$3.1 \times 10^{-5}$&$8.9 \times 10^{-5}$& $2.5 \times 10^{-4}$&$8.8 \times 10^{-4}$\\
$^{26}$Mg&$2.8 \times10^{-8}$ &$1.2 \times 10^{-6}$&$3.7 \times 10^{-6}$& $1.1 \times 10^{-5}$&$4.1 \times 10^{-5}$\\
$^{26}$Al &$1.6 \times10^{-7}$&$8.5 \times 10^{-6}$&$3.2 \times 10^{-5}$& $9.9 \times 10^{-5}$&$3.4 \times 10^{-4}$\\
$^{27}$Al &$7.8 \times10^{-7}$&$3.8 \times 10^{-5}$&$1.9 \times 10^{-4}$& $4.2 \times 10^{-4}$&$1.7 \times 10^{-3}$\\
$^{28}$Si &$7.1 \times10^{-5}$&$1.8 \times 10^{-3}$&$2.3 \times 10^{-3}$& $2.8 \times 10^{-2}$&$5.6 \times 10^{-2}$\\
$^{29}$Si &$1.7 \times10^{-6}$&$3.8 \times 10^{-5}$&$4.5 \times 10^{-5}$& $4.6 \times 10^{-4}$&$1.2 \times 10^{-3}$\\
$^{30}$Si &$8.1 \times10^{-5}$&$5.4 \times 10^{-4}$&$2.2 \times 10^{-4}$& $1.1 \times 10^{-2}$&$1.5 \times 10^{-2}$\\
$^{31}$P &$1.2 \times10^{-5}$&$1.8 \times 10^{-4}$&$3.9 \times 10^{-5}$& $3.7 \times 10^{-3}$&$4.7 \times 10^{-3}$\\
$^{32}$S &$3.9 \times10^{-3}$&$5.2 \times 10^{-4}$&$2.1 \times 10^{-4}$& $1.0 \times 10^{-2}$&$4.2 \times 10^{-3}$\\
$^{33}$S &$1.1 \times10^{-5}$&$2.2 \times 10^{-6}$&$1.6 \times 10^{-6}$& $1.5 \times 10^{-5}$&$4.4 \times 10^{-6}$\\
$^{34}$S &$8.6 \times10^{-6}$&$6.5 \times 10^{-6}$&$6.8 \times 10^{-6}$& $6.6 \times 10^{-6}$&$5.3 \times 10^{-6}$\\
$^{35}$Cl &$4.8 \times10^{-5}$&$1.1 \times 10^{-5}$&$4.7 \times 10^{-6}$& $1.3 \times 10^{-5}$&$6.5 \times 10^{-6}$\\
$^{36}$Ar &$4.3 \times10^{-6}$&$1.6 \times 10^{-5}$&$2.5 \times 10^{-5}$& $5.1 \times 10^{-6}$&$1.1 \times 10^{-5}$\\
$^{40}$Ca &$8.7 \times10^{-5}$&$5.4 \times 10^{-5}$&$3.6 \times 10^{-5}$& $4.8 \times 10^{-5}$&$3.0 \times 10^{-5}$\\
\enddata
\tablenotetext{a}{Radii from \citet{althaus_2022_ab, althaus_2023_aa}} 
\tablenotetext{b}{This simulation did not ejecta any matter so these are the abundances in the surface zone}
\tablenotetext{c}{ Composition: Carbon-Oxygen Core 25\% and 75\% Solar}
\tablenotetext{d}{ Composition: Carbon-Oxygen Core 50\% and 50\% Solar}
\tablenotetext{e}{ Composition: Oxygen-Neon Core 25\% and 75\% Solar}
\tablenotetext{f}{ Composition: Oxygen-Neon Core 50\% and 50\% Solar.}
\tablenotetext{g}{All abundances are mass fraction}
\tablenotetext{h} {The abundance of $^7$Li was less than $10^{-15}$.}
\end{deluxetable}

Figure \ref{wd1827} and Table \ref{rad1827abund} show the same data as in Figure \ref{wd1522} and Table \ref{rad1522abund} 
but for the 1827 km WD.
Figure \ref{wd2166} and Table \ref{rad2166abund} show the data for the 2166 km WD.
The gross features at both larger radii are quite similar to the 1522 km WD simulation although $^3$He is less depleted,
and $^{24}$Mg and $^{26}$Mg are more depleted.  $^{20}$Ne shows large enrichments in both
 ONe compositions but not the CO compositions.  This is because the initial $^{20}$Ne abundance is 
 significantly enriched (the values are taken from \citet{kelly_2013_aa} for both the ONe 25/75 and
 ONe 50/50 compositions).   Those values are $7.95 \times 10^{-2}$ for the ONe 25/75 mixture and
 $0.157 $ for the ONe 50/50 mixture.  The corresponding values for the CO 25/75 and CO 50/50 mixtures
 are $8.752 \times 10^{-4}$ and $5.837\times10^{-4}$, respectively. 
 The latter two values are below  solar \citep{lodders_2021_aa} because they have been mixed with ONe WD core matter. 
 
 $^{30}$Si and $^{31}$P are both enriched by about a factor of $10^3$ in the 
ONe 25/75 and ONe 50/50 simulations on all three WD radii but much less enriched in both the CO simulations.  
The enrichment of $^{30}$Si is important because NIR spectroscopy of LMCN 1968-12a, a RN in the Large
Magellanic Cloud (LMC) with a 4 yr eruption period, shows the [Si X] coronal line at 1.4309 $\mu$m extraordinarily strong
 \citep[$\sim 95$ L$_\odot$:][]{evans_2025_aa}.

The effects of WD radius and composition are complicated for the sulfur isotopes.  For the 1522 km WD, $^{33}$S is 
the most enriched of the three sulfur isotopes and there is hardly any difference in the results for either the
ONe 25/75 or ONe 50/50 compositions.  In contrast, $^{32}$S is the most enriched of the three isotopes 
on the 1827 km WD and $^{34}$S is not enriched.  The simulations on the 2166 km WD indicate
that $^{32}$S is again the most enriched and $^{34}$S is actually depleted for all four mixed compositions.

Table \ref{tcrbtable2} gives the ejected $^7$Li (assuming the ejected $^7$Be has decayed to the tabulated $^7$Li) in solar masses for the 1522 km WD.
The CO 25/75 composition ejects nearly 100 times more $^7$Li  than the ONe 25/75 composition. 
In contrast, the ONe 25/75 composition ejects more
$^{22}$Na and nearly 100 times more $^{26}$Al than the CO 25/75 composition.

Comparing the CO 50/50 to ONe 50/50 simulations on the 1522 km WD (Table \ref{rad1522abund}), the odd light isotopes are more enriched than the even light isotopes for both compositions.  $^{22}$Ne is much more depleted in the ONe simulation than the CO simulation and the intermediate mass isotopes are more enriched in the ONe simulation than the CO simulation.  Table \ref{tcrbtable2}  also shows that $^7$Li is more enriched in the CO 50/50 simulation
while both $^{22}$Na and $^{26}$Al are more enriched in the ONe 50/50 simulation.  
The sequences with 50\% WD core and 50\% solar enrichment show a larger variation when we compare the 1522 km WD to the larger radii
WDs (Figures \ref{wd1827} and \ref{wd2166}).

Table \ref{rad1522abund} (1522 km radius WD), Table \ref{rad1827abund} (1827 km radius WD),  and Table \ref{rad2166abund} (2166 km radius WD) allow
us to compare the detailed abundance predictions for most of the nuclei in our network.  Each of the three tables is for 
simulations on a different WD radius.
Because the accreting matter mixes with WD core matter, the hydrogen abundance is always sub-solar. 
In addition, the energy that drives
the explosion comes from the fusion of hydrogen to helium via either the proton-proton chain (during the majority of the accretion
phase) or the CNO cycle (during the explosive phase of the TNR). Examining these tables, we see
that the ejected hydrogen abundance ranges from a high of 0.44 (mass fraction) to a low of 0.18.  This large a range has been seen in 
spectroscopic studies of CN ejecta \citep[][and references therein]{gehrz_1998_aa, dellavalle_2020_aa, chomiuk_2021_aa} and we can expect the ejecta of TCrB to lie within this range.
As a result of the TNR, $^4$He is always enriched. Its ejecta abundance ranges from a high of 0.34 to a low of 0.21.

For the larger radii WDs,  both  $^{16}$O and $^{17}$O are considerably more enriched. In fact, $^{16}$O is more than 8\% of the ejected matter for the 2166 km WD ejecta and $^{17}$O is 5\%. Such large enrichments should be
easily observable in the spectra of TCrB and could provide an estimate of the WD mass.  
The major enrichments are for the isotopes with A $>$30 just as we saw in the 
CO 25/75 and ONe 25/75 simulations on the 1522 km WD.   The largest depletion is for $^{34}$S.

In virtually all cases our simulations predict that the abundance of  $^{13}$C exceeds that of $^{12}$C, $^{15}$N exceeds $^{14}$N, 
and  $^{17}$O exceeds $^{16}$O. These ratios are not observed in either CNe or RNe.  
While \citet{sneden_1975_aa} reported enriched $^{13}$C and $^{15}$N in the ejecta of Nova DQ Her(1934), 
their observations did not imply that the abundance of $^{13}$C exceeded that of $^{12}$C.  
\citet{banerjee_2016_aa}, using near-infrared spectroscopy of V5668 Sgr reported $^{12}$C/$^{13}$C =1.5,
and \citet{joshi_2017_aa} report a $^{12}$C/$^{13}$C ratio of 1.6 for Nova Oph 2017.
Recently, \citet{rudy_2024_aa}, using infrared observations of nova V1391 Cas,  reported $^{12}$C/$^{13}$C = 2
and not the reverse. {Determination of the $^{12}$C/$^{13}$C ratio during the imminent eruption of TCrB will be crucial. Although SyRNe are not noted for CO and dust formation during outburst, they were present - briefly - during the 2014 eruption of V745 Sco \citep{banerjee_2023_aa}. We strongly encourage infrared spectroscopy during TCrB's imminent eruption to monitor for CO formation, and to determine the $^{12}$C/$^{13}$C ratio.}

We suggest, therefore, that the $^{12}$C/$^{13}$C ratios found in this paper can be explained by mixing of the ejecta with
 material in the accretion disk that has not fallen onto the WD.  It is also possible that the ejecta in the orbital plane have
 mixed with pre-existing gas ejected from the secondary prior to the TCrB outburst \citep{williams_2008_aa}.  
 If either of these is the solution, then we predict
 that there will be different abundances measured for material in the equatorial plane of these systems when 
 compared to material ejected in the polar directions \citep[see also][and references therein]{munari_2024_aa}.   
 This is an area of discovery for JWST's integral field unit spectrographs (that can deliver high spatial resolution spectral data cubes over several square arcseconds) for the investigation of evolved CNe and RNe.
 
 {Our analyses of the IFU data will be guided by recent IFU studies of nova shells done with ground based instruments.
 In particular we mention studies of QU Vul \citep{santamaria_2022_aa, santamaria_2022_ab}, T Aur, Hr Del, 
 DQ Her \citep{santamaria_2022_aa}, RR Pic \citep{celedon_2024_aa}, V1425 Aql \citep{celedon_2025_aa}, and  T Pyx \citep{izzo_2024_aa, guerrero_2025_aa}.  They
 report that these CN shells are prolate spheroids and the ejected mass is in some cases factors of 10 larger
 then predictions.}  
 
 The last three isotopes that we tabulate in Tables \ref{rad1522abund}, \ref{rad1827abund} , and \ref{rad2166abund} are
 $^{35}$Cl, $^{36}$Ar, and $^{40}$Ca.  While the isotopes in the network and the network solver extend
 to more massive isotopes, it is normally $^{40}$Ca, and more massive isotopes,  that show no effects of the TNR
 and we see that in both the tables and the figures.  
 \citet{jose_2007_aa} report the same result for solar metallicity studies but find that more massive isotopes are 
 extremely enriched in
 low metallicity simulations. In contrast, $^{35}$Cl is enriched in all mixed compositions on the 1522 km and 1827 km WDs, 
 but only slightly enriched in the ONe compositions on the 2166 km WD.  Again, this is a result of the higher peak temperatures
 on the smaller WDs.  While $^{36}$Ar is enriched in the ONe 25/75 simulation on the 1522 km WD, it is depleted in the
 other three mixtures on this WD. It is depleted for all 4 mixed compositions on the 1827 km and 2166 km WDs. 
 
Finally, we discuss two radioactive nuclei: $^{22}$Na and $^{26}$Al.  Detailed predictions for them can be found in \citet{starrfield_2024_aa} so we only
summarize the new simulations here.  Examining all the tables, we find that the maximum amount of $^{22}$Na ejected is $1.8 \times 10^{-11}$ M$_\odot$
for the ONe 50/50 simulation on the 1522 km WD. This compares to a value of $1.43 \times 10^{-8}$ M$_\odot$ in the CN simulation that
was done for a 1.35 M$_\odot$ WD with an ONe 50/50 composition \citep{starrfield_2024_aa}. The ratio of the evolution times to TNR is
$\sim$ 750 suggesting over the lifespan of TCrB that it can eject comparable amounts of $^{22}$Na as a CN.  Performing the same
comparison for $^{26}$Al, we find that the peak value for this isotope is $3.0\times 10^{-11}$M$_\odot$ on the 2166 km WD with the
ONe 50/50 composition.  This value compares to $1.87 \times 10^{-8}$M$_\odot$ on the 1.35M$_\odot$ simulation with the ONe 50/50
composition \citep{starrfield_2024_aa}. The ratio is 620 so that repeated outbursts on TCrB can equal the amount ejected by a CN.  

\needspace{2 \baselineskip}
\bigskip
\section{Evolution of the White Dwarf After the End of the Explosive Phase}
\label{remnant}

As noted in Section \ref{intro}, a unique feature of the outburst of TCrB was the secondary
maximum that occurred about 100 days after the 1946 explosion. 
In order to investigate this behavior,  we
use NOVA to follow the evolution of just the WD after the explosive phase of the outburst.  

For reasons stated in Section \ref{novacode}, NOVA cannot follow the evolution of the material that has reached escape velocity, and is
optically thin, for the 100 or more days necessary to test the hypothesis of \citet{munari_2023_aa}. This is because the density of the
rapidly expanding matter falls below the low density limit of the EOS tables and we are forced to end the
evolution.

\begin{figure}[htb!]
\gridline{\fig{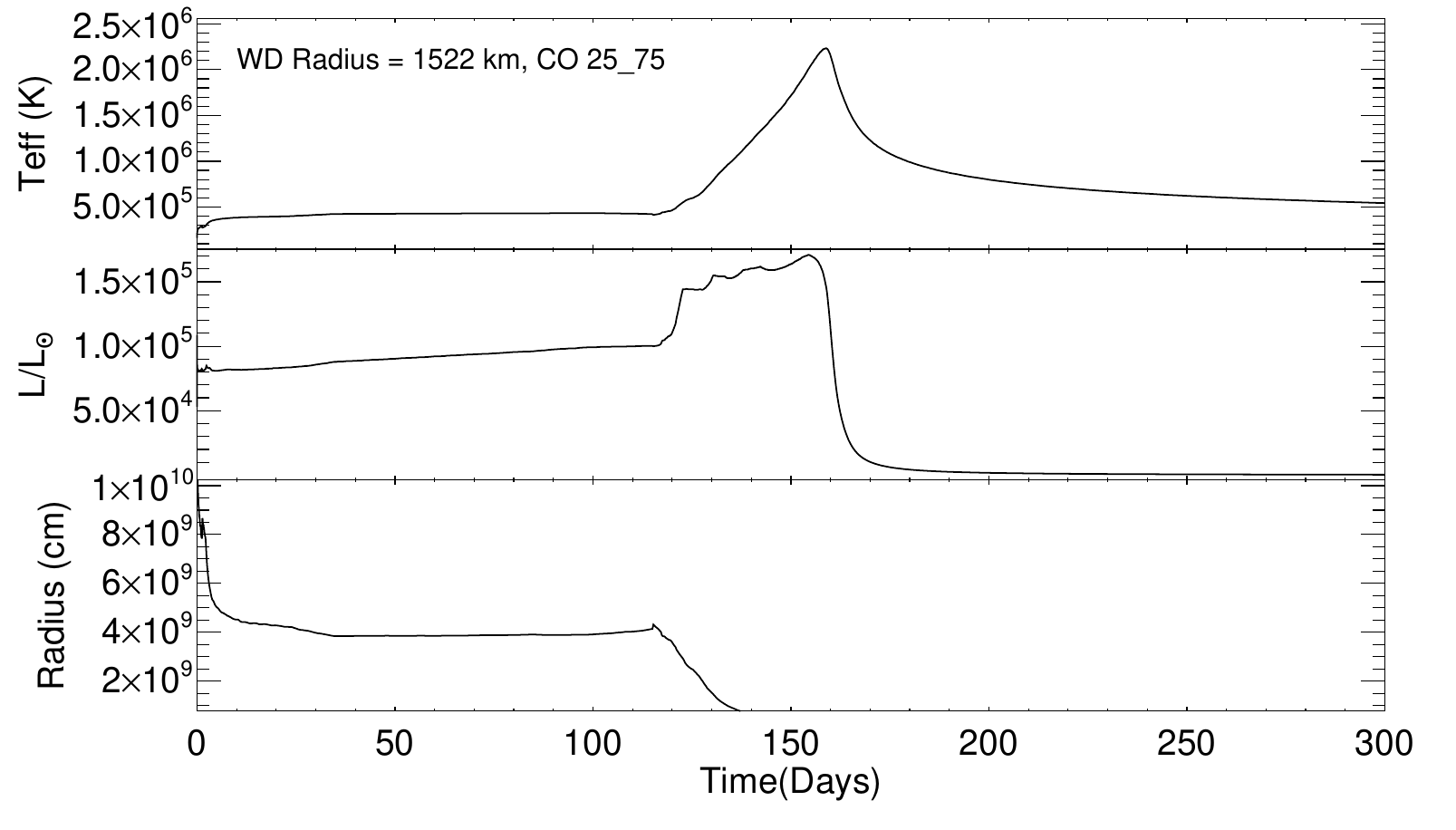}{0.5 \textwidth}{}
            \fig{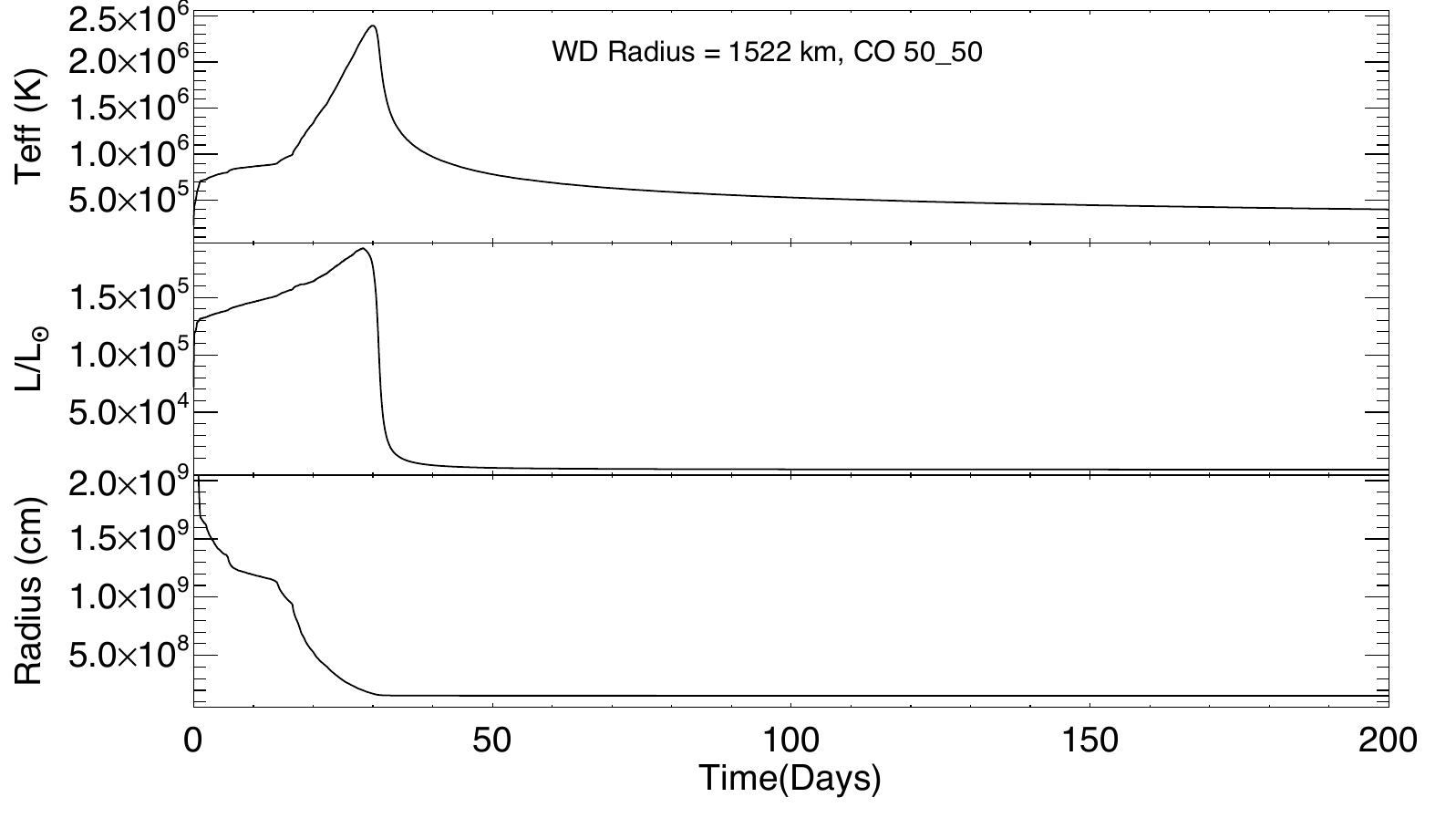}{0.5\textwidth}{}}
\gridline{\fig{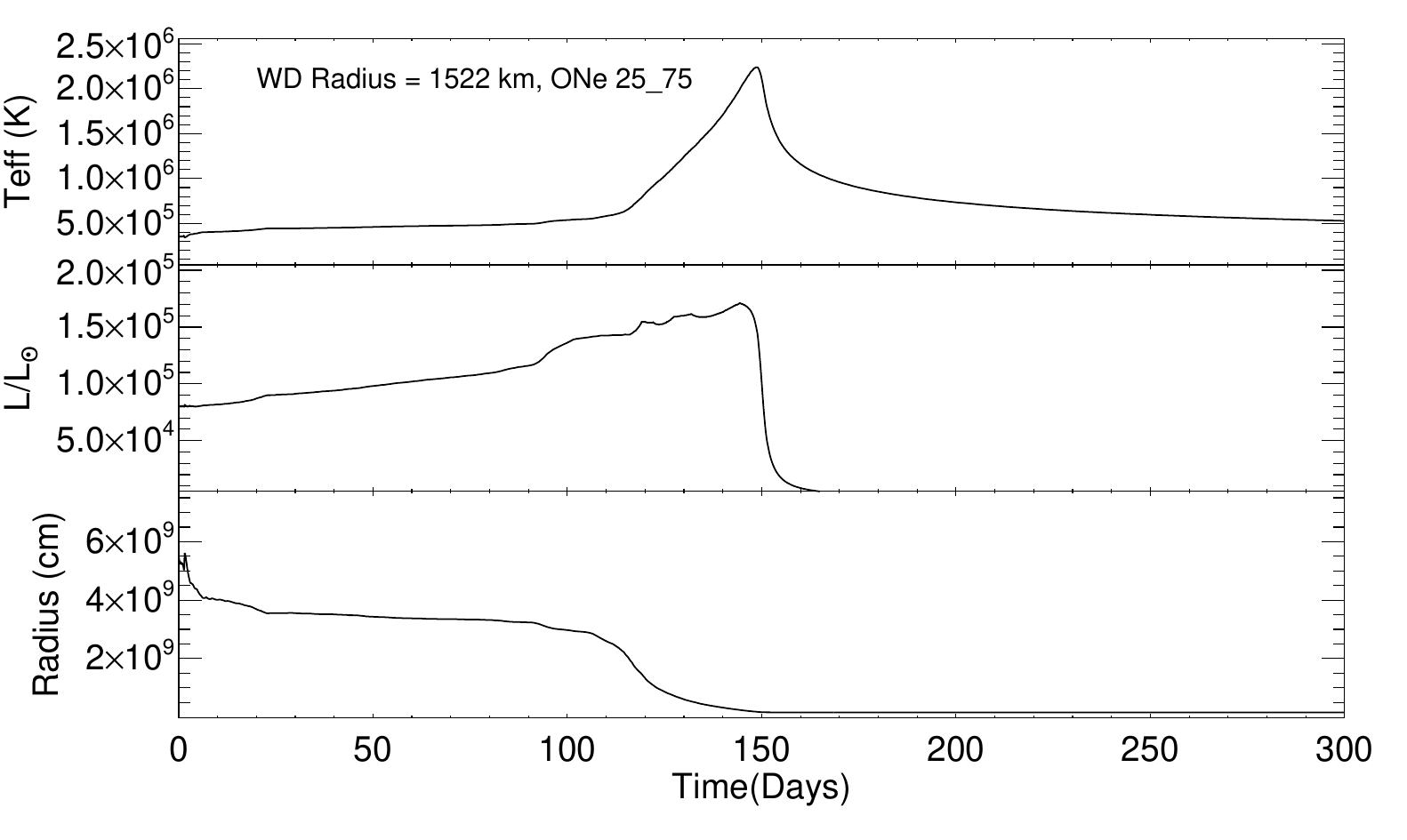}{0.5\textwidth}{}
            \fig{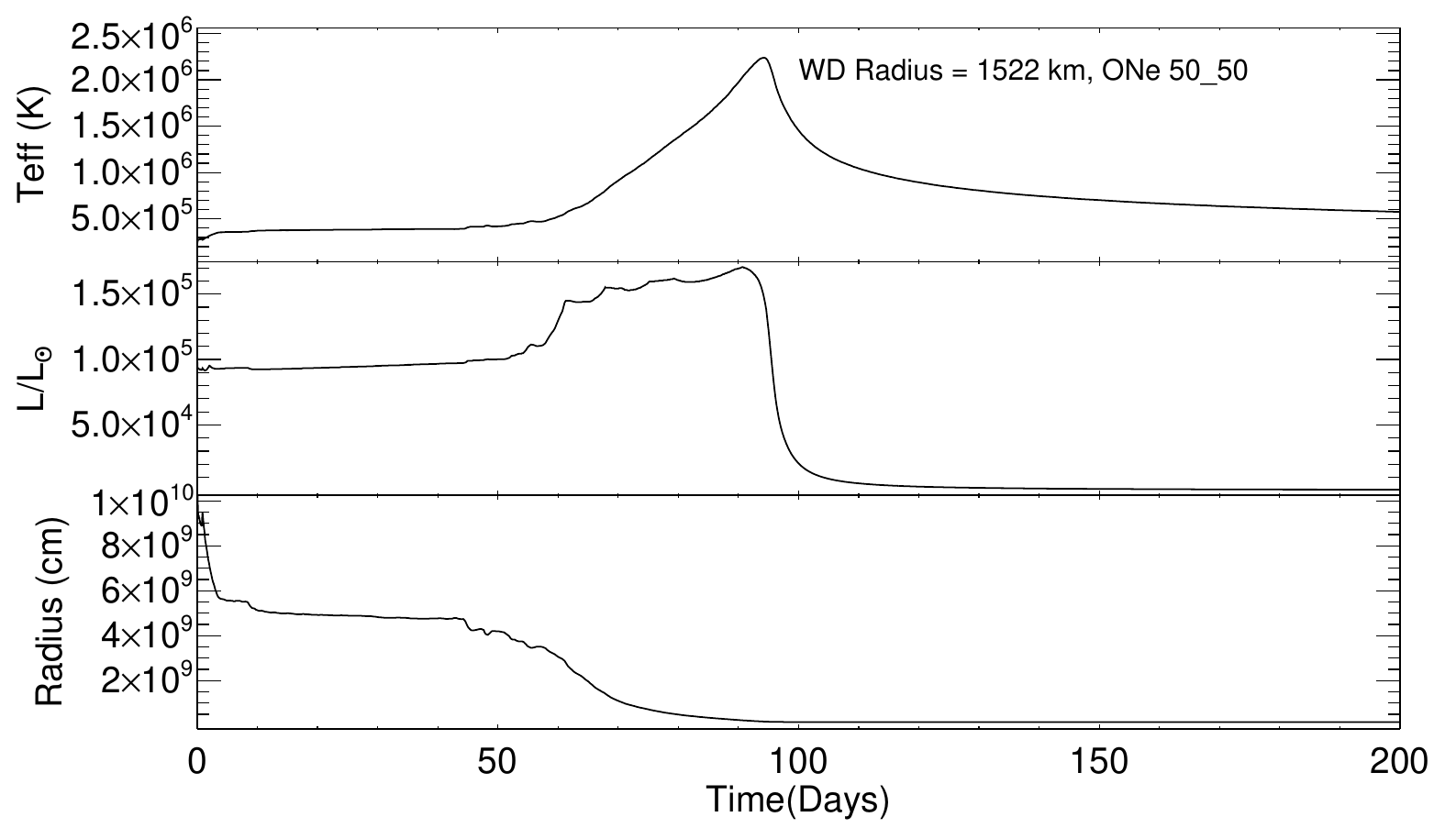}{0.5\textwidth}{}}
\caption{1522 km WD: These 4 plots, each with 3 panels, show the evolution of the effective temperature, T$_{\rm eff }$ (K), the
luminosity, L/L$_\odot$, and the radius (cm) of the surface layers of the remnant WD after the removal of the ejected matter. They are all for the
same initial WD radius but for each of the 4 mixed compositions.  The particular composition used for each plot
is given on the plot.   They show that the surface
conditions are relatively constant for a short time (days to weeks depending on the conditions) 
and then as the hydrogen burns out the radius
begins a decline, the luminosity and effective temperature increase, and then decline. See the text for more details on the
evolution.}
\label{figure1522}
\end{figure}

However, we can
remove the escaping low density material (mass zones) from the numerical mesh and apply new surface boundary conditions
that apply the minimum numerical pressure to the new surface zones that
prevent the underlying zones from accelerating and escaping from the WD.  
This procedure ensures that their velocities are virtually unchanged from before the 
mass removal until just after.  We then follow the evolution of the remnant WD until the remaining hydrogen burns out and the
WD evolves back to quiescence.  In this section, we report on the results of applying this technique to
all the mixed composition simulations discussed in Section \ref{results}.   
 
 \begin{deluxetable}{@{}lccccccc}
\tabletypesize{\small}
\tablecaption{Evolutionary results for  the White Dwarf after Removal of the Ejected Matter\tablenotemark{a}\label{tcrbtable6}}
\tablewidth{0pt}
\tablecolumns{6}
 \tablehead{ \colhead{Composition:} &
 \colhead{CO 25/75}\tablenotemark{b} &
\colhead{CO 50/50}\tablenotemark{c} &
 \colhead{ONe 25/75}\tablenotemark{d} &
 \colhead{ONe 50/50\tablenotemark{e}}}
 \startdata
{\bf WD Radius = 1522 km\tablenotemark{}}&&&&\\
 \hline
T$_{\rm eff~peak}$($10^6$K)&2.23&2.39&2.24&2.24\\
Time to T$_{\rm eff~peak}$(days)\tablenotemark{}&158.8&29.9&148.5&94.1\\
L$_{\rm peak}$/L$_\odot$ ($10^5$)&1.71&1.93&1.71&1.71\\
Time to L$_{\rm peak}$(days)&154.5&28.3&144.2&90.6\\
M$_{\rm ej}$($10^{-8}$M$_{\odot}$)&$5.6$&$42.0$&$4.6$&5.3\\
RE: (M$_{\rm acc}$ -  M$_{\rm ej}$)/M$_{\rm acc}$&0.94&0.56&0.95&0.94\\
\hline
{\bf WD Radius = 1827 km\tablenotemark{}}&&&&&&\\
\hline
T$_{\rm eff~peak}$($10^6$K)&1.99&2.10&1.96&2.03\\
Time to T$_{\rm eff~peak}$(days)\tablenotemark{}&247.9&63.6&253.4&101.5\\
L$_{\rm peak}$/L$_\odot$ ($10^5$)&1.59&1.79&1.58&1.66\\
Time to L$_{\rm peak}$(days)&239.0&60.7&244.1&97.0\\
M$_{\rm ej}$($10^{-8}$M$_{\odot}$)&2.6&35.0&1.5&11.0\\
RE: (M$_{\rm acc}$ -  M$_{\rm ej}$)/M$_{\rm acc}$&0.98&0.72&0.99&0.92\\
\hline
{\bf WD Radius = 2166 km\tablenotemark{}}&&&&&&\\
\hline
T$_{\rm eff~peak}$($10^6$K)&1.83&1.88&1.79&1.82\\
Time to T$_{\rm eff~peak}$(days)\tablenotemark{}&318.3&128.2&326.8&175.9\\
L$_{\rm peak}$/L$_\odot$ ($10^5$)&1.56&1.70&1.51&1.56\\
Time to L$_{\rm peak}$(days)&312.5&120.0&319.7&162.0\\
M$_{\rm ej}$($10^{-8}$M$_{\odot}$)&$6.0$&$34.9$&$3.2$&1.2\\
RE: (M$_{\rm acc}$ -  M$_{\rm ej}$)/M$_{\rm acc}$&0.97&0.82&0.98&0.99\\
\enddata
\tablenotetext{a}{Radii from \citet{althaus_2022_ab, althaus_2023_aa}}
\tablenotetext{b}{Composition: Carbon-Oxygen Core 25\% and 75\% Solar}
\tablenotetext{c}{Composition: Carbon-Oxygen Core 50\% and 50\% Solar}
\tablenotetext{d}{Composition: Oxygen-Neon Core 25\% and 75\% Solar}
\tablenotetext{e}{Composition: Oxygen-Neon Core 50\% and 50\% Solar.}
\end{deluxetable}

We summarize those results in Table \ref{tcrbtable6}. Again, the rows are organized by the initial WD radius. For
each radius, we give the peak effective temperature, T$_{\rm eff~peak}$, the time in days, after the escaping layers
are removed, at which that
temperature is reached, Time to T$_{\rm eff~peak}$, the peak luminosity in solar units, L$_{\rm peak}$/L$_\odot$,
and the time to reach that luminosity in days.  We follow with both the ejected mass and the RE copied from
Table \ref {tcrbtable2} in order to show that the amount of ejected mass does affect the time it takes for the WD to return
to quiescence.  Table \ref{tcrbtable6} shows that the only simulations with evolution times close to what
\citet{munari_2023_aa} predicts are the ONe 50/50 simulations on the 1522 km and 1827 km WDs.

\begin{figure}[htb!]
\gridline{\fig{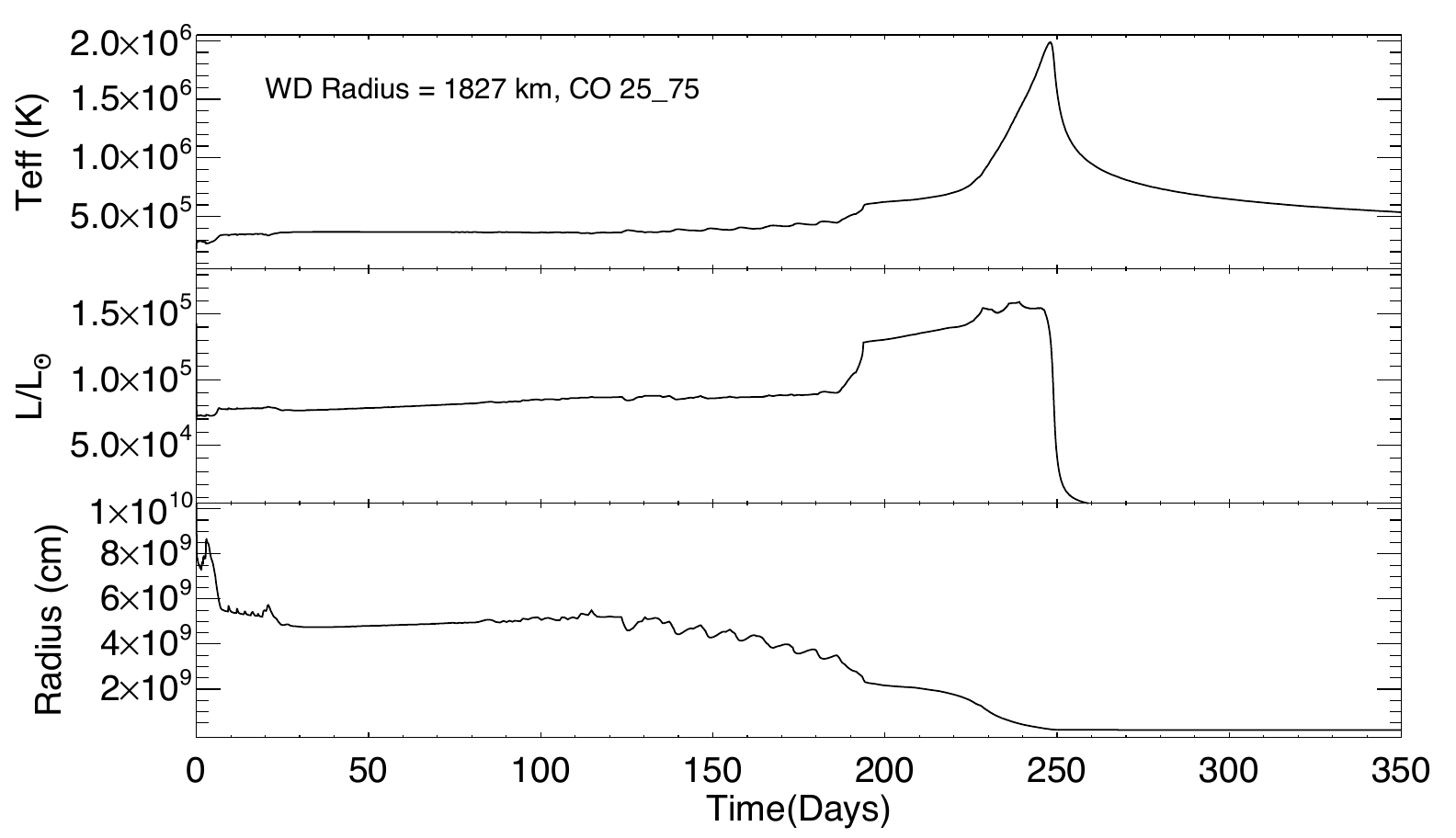}{0.5\textwidth}{}
            \fig{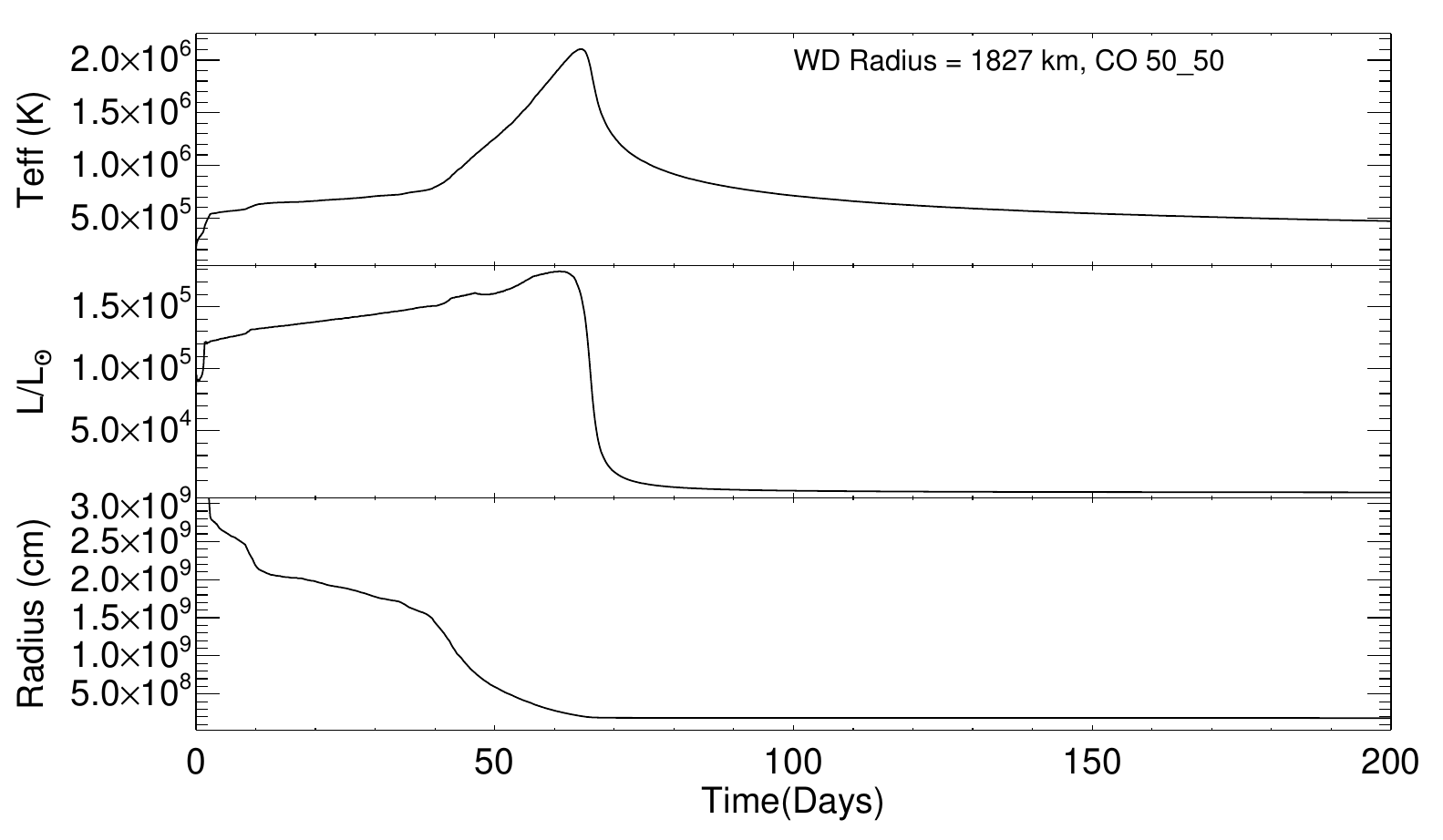}{0.5\textwidth}{}}
\gridline{\fig{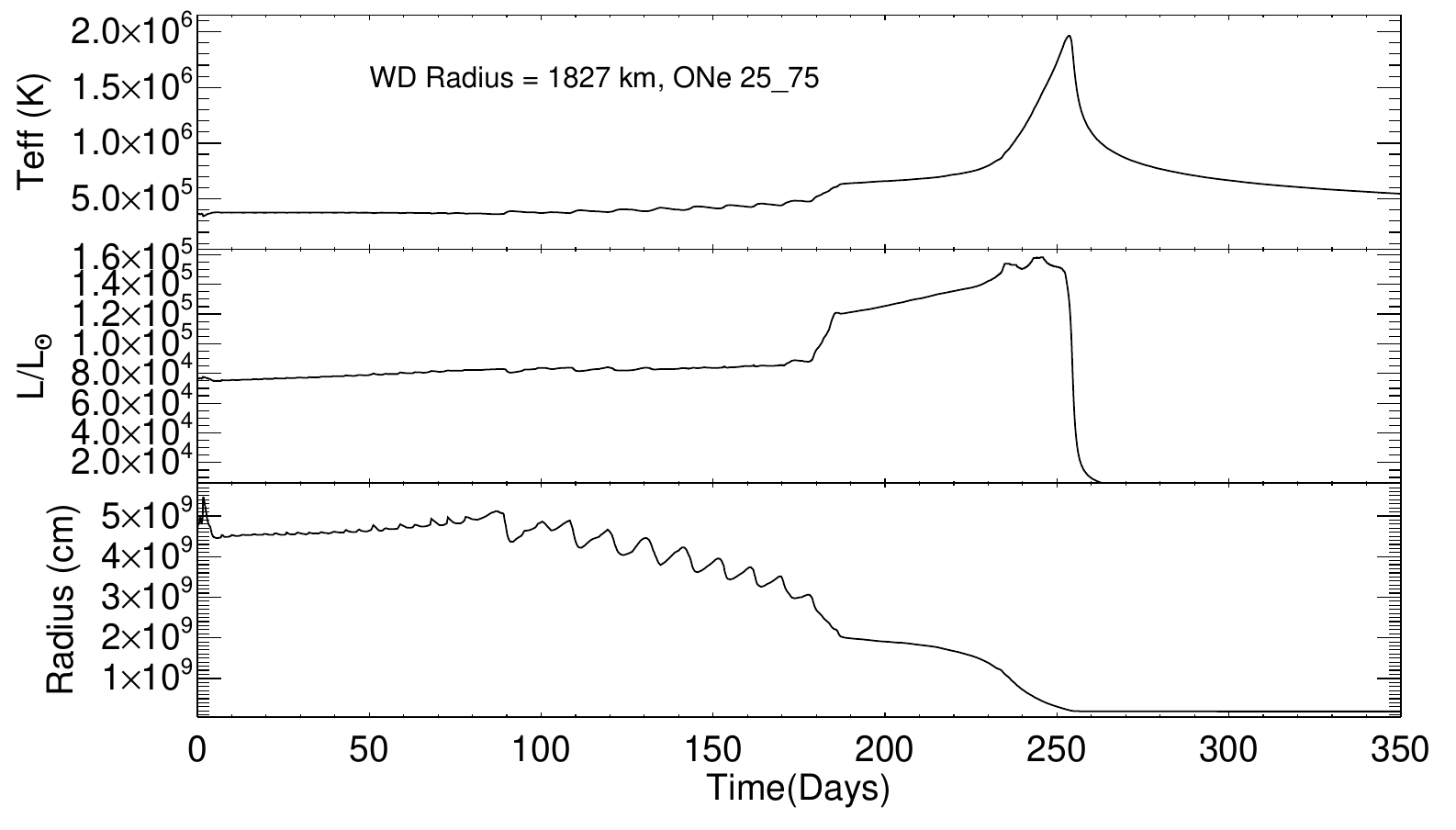}{0.5\textwidth}{}
            \fig{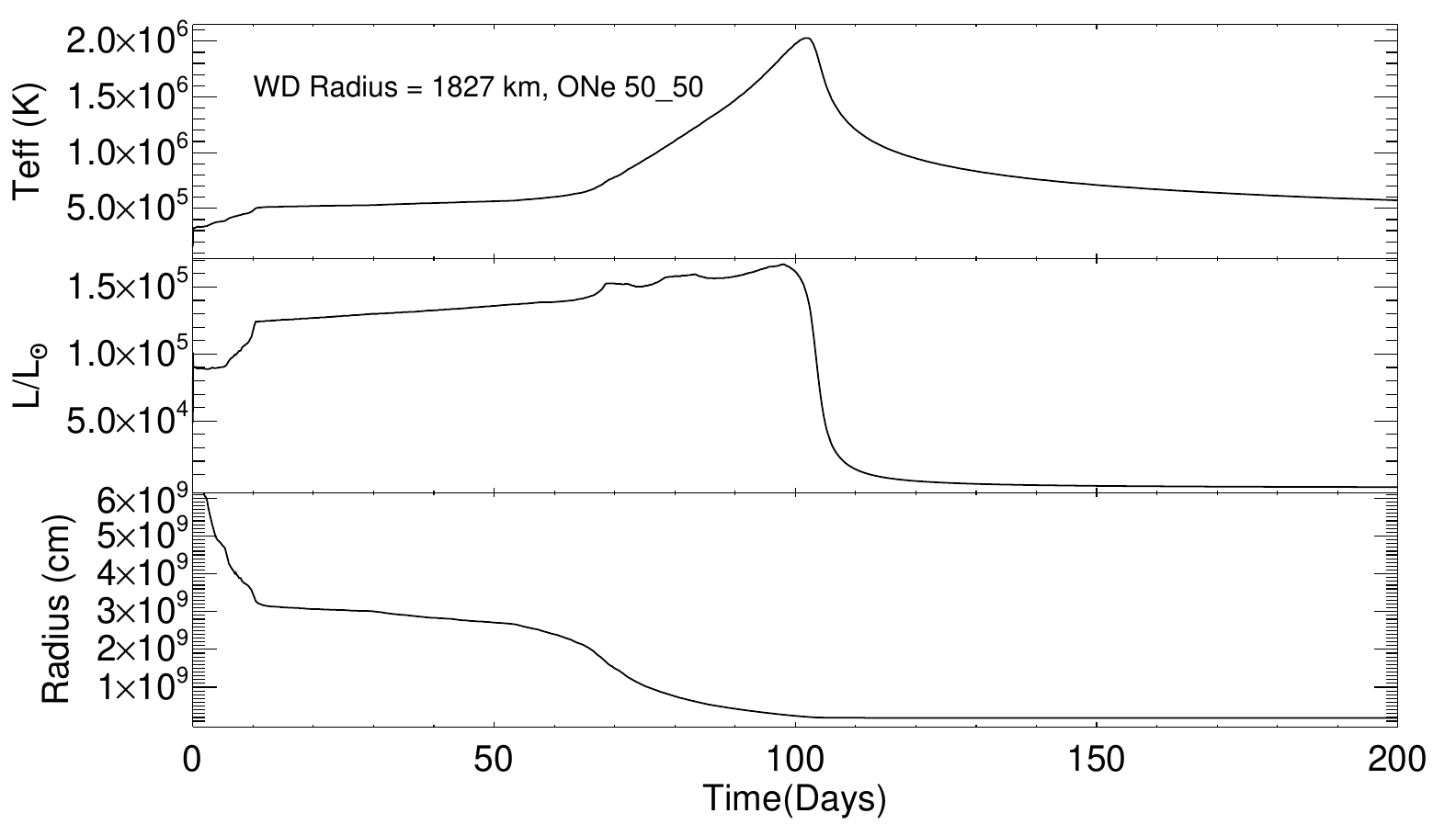}{0.5\textwidth}{}}
\caption{1827 km WD: Otherwise the same plot as for Figure \ref{figure1522}. }
\label{figure1827}
\end{figure}

\begin{figure}[htb!]
\gridline{\fig{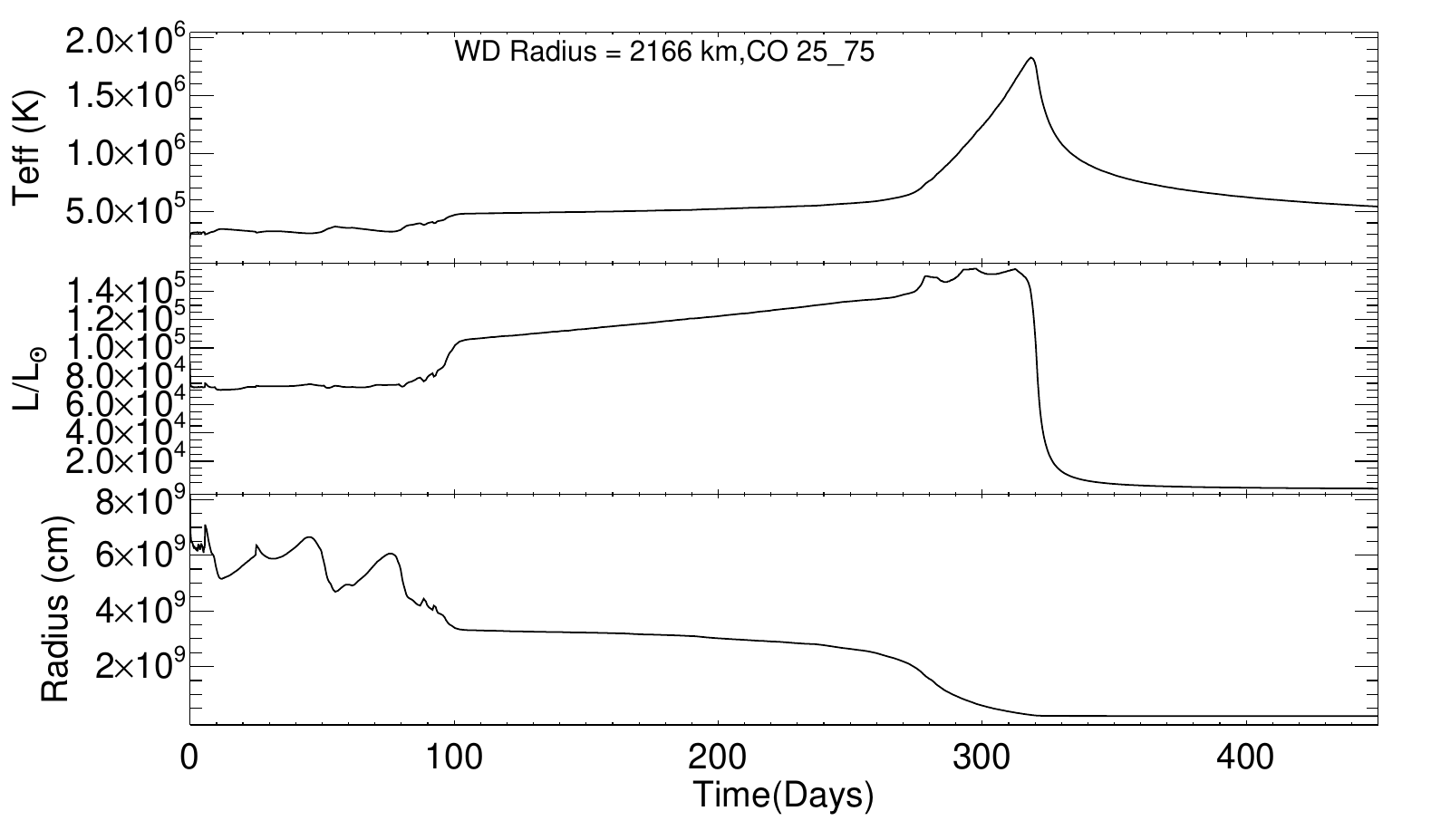}{0.5\textwidth}{}
            \fig{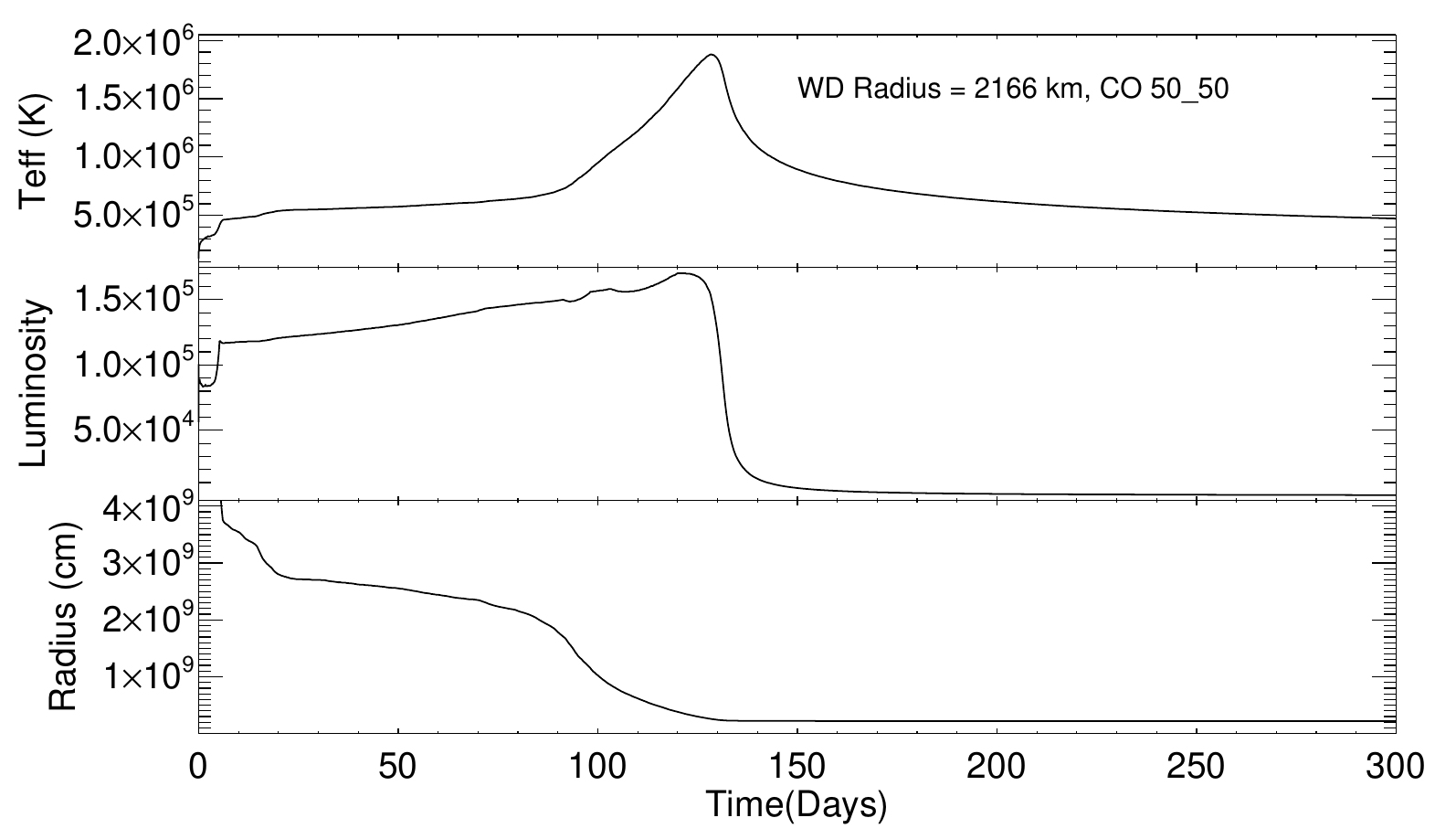}{0.5\textwidth}{}}
\gridline{\fig{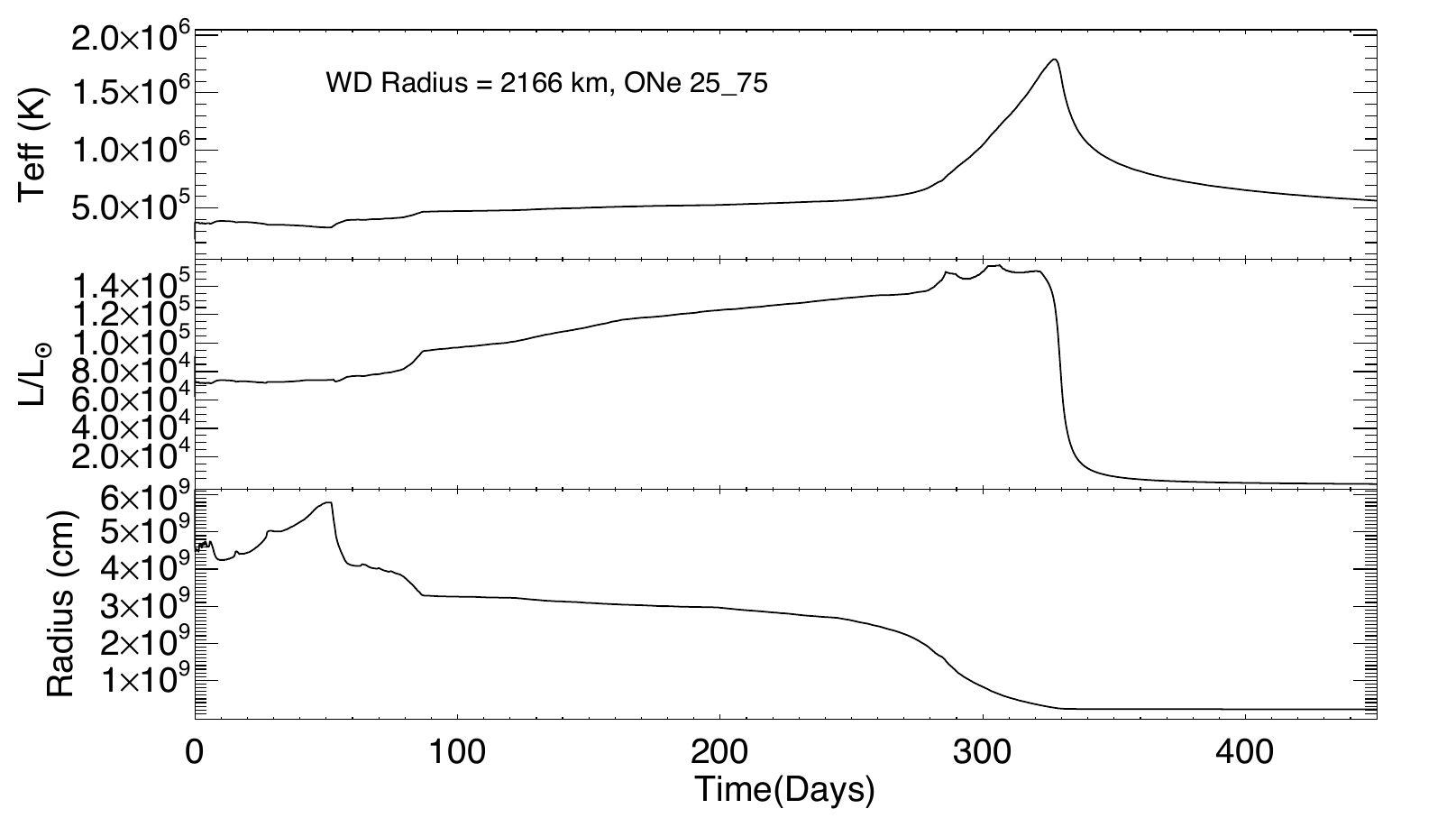}{0.5\textwidth}{(c)}
            \fig{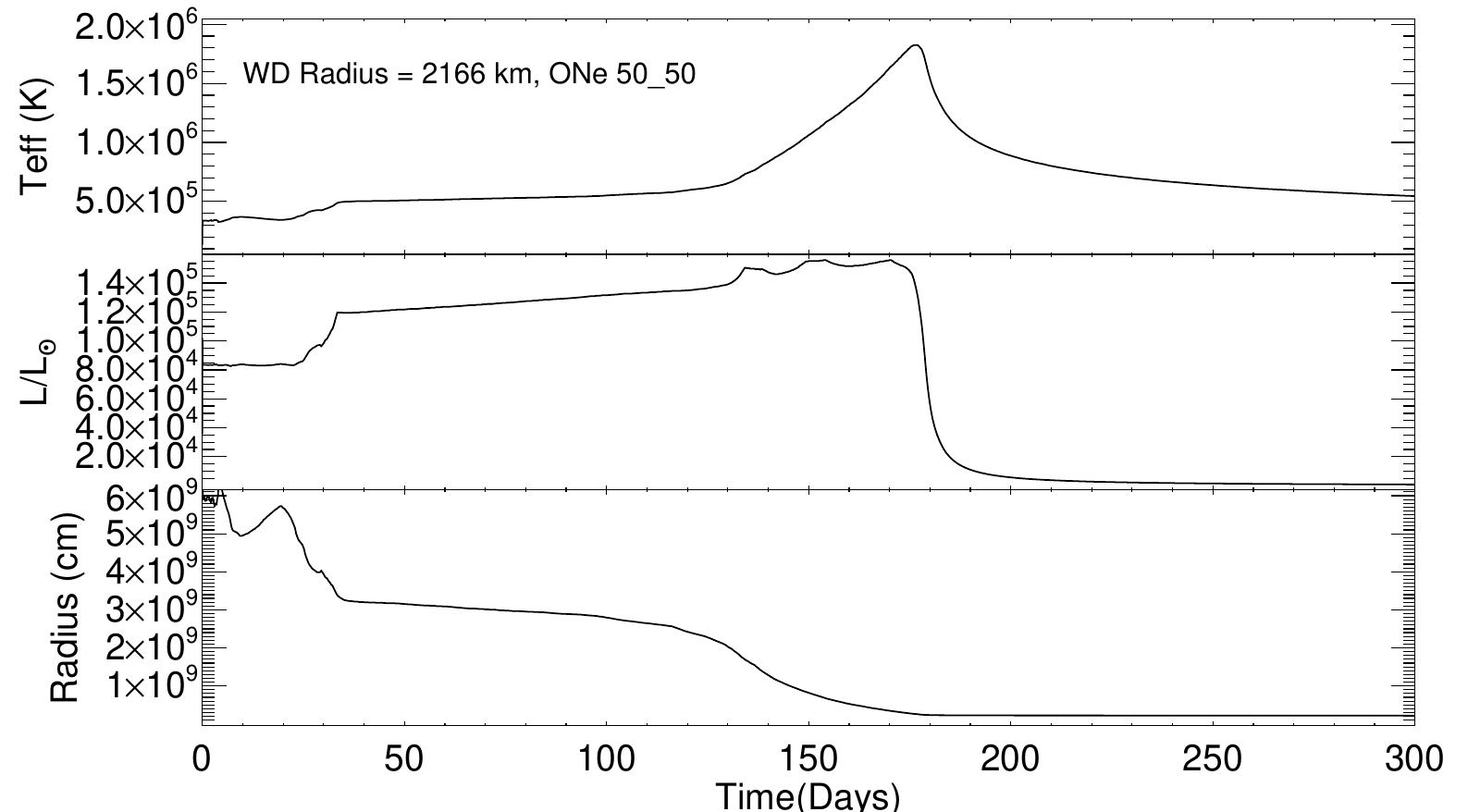}{0.5\textwidth}{(d)}}
\caption{2166 km WD: The same 
plot as for Figure \ref{figure1522}.}
\label{figure2166}
\end{figure}

In Figures \ref{figure1522}, \ref{figure1827}, and \ref{figure2166}, we show the evolution of all the simulations
from the time that the ejected layers are removed until the return to quiescence of the WD. Each of the 
subplots consists of 3 panels.  The top panel shows the evolution of the effective temperature,
the middle panel shows the evolution of the luminosity in solar units, and the bottom panel
gives the value of the radius of the surface layers of the remnant WD at that time.  The zero point in time is when the ejected matter is removed.

In all cases, the initial WD effective temperature exceeds $\sim 2 \times 10^5$ K.  The effective temperature stays
at that value until the radii of the surface layers begin their decline, 
as the remaining hydrogen begins to burn out. It then starts to increase, reaching in some cases,
values of $\sim 2.5 \times 10^6$ K.  At the same time, the luminosity, which has stayed 
roughly constant or is gradually increasing, rapidly climbs to values of $\sim1.5 \times 10^5$ L/L$_\odot$.
This evolutionary behavior is the result of the hydrogen, remaining in the envelope, undergoing nuclear burning 
at high rates. As the hydrogen burns out, the WD outer layers begin to shrink on the Kelvin-Helmholtz
time scale, the luminosity increases, and the effective temperature increases. This mechanism
was applied to the turn-off of V1974 Cyg \citep{krautter_1996_aa}.

 \begin{deluxetable}{@{}ccccccccc}
\tabletypesize{\small}
\tablecaption{Detailed Evolution of the 1522 km CO 25/75 Simulation\tablenotemark{}\label{tcrbtable7}}
\tablewidth{0pt}
\tablecolumns{9}
 \tablehead{ \colhead{Age} &
 \colhead{L/L$_\odot$}\tablenotemark{} &
\colhead{T$\rm_{eff}$}\tablenotemark{} &
 \colhead{Radius}\tablenotemark{a} &
 \colhead{T$\rm_{max}$}\tablenotemark{} &
 \colhead{$\epsilon\rm_{nuc}$}\tablenotemark{} &
 \colhead{V}\tablenotemark{} &
 \colhead{X}\tablenotemark{} &
  \colhead{Y\tablenotemark{}}}
 \startdata
 \hline
 (days)&($10^5$)&($10^5$K)&($10^8$cm)&($10^8$K)&($10^{12}$)\tablenotemark{b}&(km s$^{-1}$)&MF\tablenotemark{c}&MF\tablenotemark{c}\\
 \hline
0.0&1.1&1.9&214.&1.2&2.4&+30.&0.37&0.37\\
0.04&0.9&2.2&145.&1.2&2.2&-9.0&0.36&0.38\\
0.45&0.8&2.7&91.&1.2&2.7&-0.4&0.35&0.39\\
1.3&0.8&2.9&79.&1.1&2.5&-0.2&0.35&0.39\\
5.4&0.8&3.7&50.&1.2&2.6&-0.03&0.33&0.41\\
11.3&0.8&3.9&44.&1.3&2.8&-0.008&0.32&0.42\\
21.9&0.8&4.0&42.&1.3&2.9&-0.005&0.29&0.44\\
40.3&0.9&4.3&38.&1.3&3.0&+0.002&0.24&0.49\\
63.5&0.9&4.3&39.&1.3&3.2&-0.0003&0.18&0.55\\
80.5&1.0&4.3&39.&1.3&3.3&+0.0004&0.14&0.60\\
99.4&1.0&4.3&39.&1.4&3.5&+0.002&0.08&0.66\\
121.1&1.2&5.0&33.&1.5&3.3&-0.10&0.03&0.71\\
128.4&1.4&7.0&18.&1.8&1.8&-0.03&$8.7\times10^{-4}$&0.74\\
132.8&1.5&9.1&11.&2.1&0.02&-0.01&$1.4\times10^{-8}$&0.74\\
136.1&1.5&10.5&8.4&2.2&0.02&-0.008&$1.6\times10^{-8}$&0.74\\
139.8&1.6&12.1&6.4&2.3&0.03&-0.005&$1.4\times10^{-8}$&0.74\\
142.3&1.6&13.3&5.3&2.3&0.04&-0.004&$1.2\times10^{-8}$&0.74\\
146.3&1.6&15.2&4.0&2.4&0.09&-0.003&$1.1\times10^{-8}$&0.74\\
150.9&1.6&18.7&3.0&2.5&0.14\tablenotemark{\bf{d}}&-0.002&$1.4\times10^{-8}$&0.74\\
154.4&1.7&20.0&2.4&2.6&0.15&-0.002&$1.2\times10^{-8}$&0.74\\
157.8&1.6&22.1&1.9&2.6&0.12&-0.001&$8.3\times10^{-9}$&0.74\\
161.3&0.7&19.5&1.6&2.4&0.05&-0.0003&$2.6\times10^{-9}$&0.74\\
166.1&0.2&14.3&1.6&2.1&0.002&-0.00005&$3.6\times10^{-11}$&0.74\\
172.0&0.08&11.6&1.6&1.8&0.0002&-0.00001&$2.7\times10^{-11}$&0.74\\
187.2&0.03&9.0&1.6&1.5&0.0&0.0&$2.6\times10^{-11}$&0.74\\
\enddata
\tablenotetext{a}{The radius of the surface layers of the remnant WD}
\tablenotetext{b}{erg gm$^{-1}$ s$^{-1}$}
\tablenotetext{c}{MF = Mass fraction in the remnant accreted layers since the envelope is completely convective.}
\tablenotetext{d}{The triple $\alpha$ reaction is becoming important in the deepest accreted zones.}
\tablenotetext{}{}
\end{deluxetable}

We follow in detail the evolution of the CO 25/75 sequence on the 1522 km WD
to demonstrate this evolution (see Table \ref{tcrbtable7}).  
The first column in Table \ref{tcrbtable7} is the age in days since the ejected matter was removed. 
 Next, we tabulate the luminosity of the WD
in units of $10^5$ L/L$_\odot$, the effective temperature in units of $10^5$ K, and the outer radius of the WD surface layers
in units of $10^8$ cm. These columns are followed by the peak temperature in the
nuclear burning envelope, T$ \rm_{max}$ (in units of $10^8$ K), the rate of nuclear
energy generation in the same mass zone, $\epsilon \rm_{nuc}$ (in units of $10^{12}$ 
erg gm$^{-1}$ s$^{-1}$), the velocity of the surface zone in km s$^{-1}$, and the
mass fractions of X and Y.

The first row, with an age of 0.0, shows the conditions in the WD at the first time
step after the removal of the ejected matter.  The irregular time intervals in the rows
in this table result from NOVA printing detailed evolutionary
properties only every 100 time steps.  

The velocity is initially positive, but by the second row (after nearly an hour),
it has turned negative and the outer radius of the WD begins to contract.  Over the first 100 days,
the evolution proceeds with only small changes in most of the quantities.
The radii of the surface layers, however, have shrunk from $\sim 10^{10}$ cm to $\sim 4 \times 10^{9}$ cm 
and the $^1$H abundance has decreased from 0.37 to 0.08.  Over the same interval, the $^4$He
abundance has increased from 0.37 to almost 0.74.  The entire accreted envelope is
convective, so these abundances are for the entire envelope, not just the
surface zones.

 Over the next $\sim$ 40 days, $^1$H burns out in the entire envelope and the
contraction accelerates.  The luminosity reaches its maximum value of 
$1.71 \times 10^5$ L$_\odot$ on about day 154 while peak T$\rm_ {eff}$ ($2.2 \times 10^6$ K) is reached
about 4 days later.  Over this same time, the radii of the surface layers have decreased to values only
a few times that of the quiescent WD.  The contraction has caused the temperature
to rise in the deepest accreted layers to almost double the values that it began with
100 days before. The rise in temperature has accelerated the depletion of hydrogen
throughout the envelope, and the triple-$\alpha$ reaction sequence has begun in the
deepest layers with $^4$He being converted to $^{12}$C.  This is not visible in the composition near
the outer layers because the layers at depth are becoming radiative and are no longer
mixing with the top of the WD.  Once the peak values have been reached, it takes only another 20 days for the
WD to contract to  quiescence. 

Upon examining the simulations, it is apparent that we bracket the proposed times
predicted by \citet{munari_2023_ab} for the cause of the secondary maximum in TCrB. However,
our simulations are extremely bright and significantly hotter than in his proposal.   As shown in our
studies of CNe \citep{starrfield_2020_aa, starrfield_2024_aa}, the WD temperatures
and luminosities are mass dependent and lower mass WDs would be cooler and fainter but the estimates of the mass of the WD in TCrB seem
reasonable.  

{One possible explanation for this difference is provided at the end of \citet{munari_2023_ab}
where he states: ``The best fit is reached for eta=0.8, but similarly good fits to II-Max can be achieved by trading a lower eta for a higher temperature and/or radius of the WD''.  In an email (Munari 2025, private communication) he continues: ``where eta is the fraction of the radiation arriving from the WD on the RG that is locally absorbed and re-thermalized. So the agreement with your models can be broader. 
My intention was to provide a simple geometrical interpretation of the secondary maximum, rather than fixing accurately the temperature/radius
of the cooling WD.''  (II-Max is the second maximum which is the subject of \citet{munari_2023_ab})}

Another possibility is that when the outer expanding layers are removed, 
the ``new'' surface zones are more massive than those that have been removed.  This is because
the mass of the zones decreases outward.   It is possible that, if we were able to 
rezone NOVA at this stage, we could evolve the WDs with much smaller outer mass zones that had a lower peak
temperature in the last phases of evolution.  Such a possibility is under investigation and beyond the scope of this paper.

\section{Discussion}
\label{discuss}

This paper continues and expands on a series of studies we have been doing for RNe in general and TCrB in 
particular \citep{starrfield_1988_aa, starrfield_2023_aa,
starrfield_2024_bb, Starrfield_2024_cc}.  While we can predict the behavior of TCrB after the TNR, predicting the time when it goes into outburst 
can only be done with a broad brush. Within the next 2 years seems likely, but not guaranteed.  When it does explode, it is expected to
reach V $\sim$2 mag, making it one of the brightest novae in the northern skies since Nova Cyg 1975.  However, that is not the entire
story; because it is so close, $\ltsimeq 1$~kpc,  our simulations of the evolution of the WD returning to quiescence after the outburst, suggest that it will become the brightest X-ray source of any recent RN or CN.  

{In addition, although this work is 1D, the actual system of a WD exploding inside the wind and atmosphere of a red giant
is an inherently 3D phenomenon.  This is clear from imaging observations of RS Oph  \citep{ribeiro_2009_aa}, V407 Cyg
\citep{giroletti_2020_aa}, Chandra spectroscopy of RS Oph \citep{drake_2009_aa}, and both CHANDRA and near-infrared spectroscopic
observations of V745 Sco \citep{drake_2016_aa, banerjee_2014_aa}.
A detailed discussion of the 3D behavior of SyRNe can be found in \citet{munari_2024_aa} and we refer the reader to
that publication.  One important result of that interaction is the high temperatures measured for the colliding gases which
in many cases exceed $10^{7}$ K \citep[\emph{e.g.,}][]{drake_2016_aa, evans_2025_aa}.
Moreover, the strong shock and short distance of TCrB imply that it will be a strong
source of VHE $\gamma$-rays \citep{tatischeff_2007_aa}.}

The results for the WD evolving to quiescence are given in Table \ref{tcrbtable6}.
In all cases, peak luminosity exceeds $\sim 1.5 \times 10^5$ L$_\odot$; and in most cases, peak T$_{\rm eff}$ 
reaches $\sim 2 \times 10^6$ K.  The results for the early phases of the outburst are given in Table \ref{tcrbtable2} and in all
cases peak T$_{\rm eff}$ exceeds $10^6$ K and the luminosity exceeds $10^4$ L$_\odot$. We predict, therefore, two episodes
of X-ray emission: a ``flash'' near or at maximum, and one about the time of the secondary maximum.  This continues our earlier
predictions of an X-ray flash which has been detected in Nova YZ Reticuli by eROSITA  \citep{konig_2022_aa}.
{Contemporaneous with the first X-ray flash is a UV flash which ionizes the red giant wind ahead of the
shock from the explosion \citep{shore_1996_aa, shore_1998_aa, munari_2024_aa}. The UV flash is
evidenced by the presence of narrow emission sitting on top of the broad emission lines.
These sharp lines disappear as the shock penetrates the entire wind \citep{munari_2024_aa}.}

The cause of these high luminosities and
effective temperatures is that convection has transported large amounts of the positron-decaying nuclei to the
surface and their decays have resulted in the energy generation in those layers exceeding  $10^{12}$ erg gm$^{-1}$ s$^{-1}$.
In some of the simulations, however, we have extrapolated the EOS to values below the low density limits in the tables and have evolved the simulation until the outer zones reached radii of  $\sim 10^{13}$ cm.  In none of these cases was significantly more material ejected (one or two mass zones at a maximum) before we ended the evolution. We examined the mass zones just below those that were escaping and found velocities that were far less than the escape velocity at that radius.

The results reported in this paper, along with our previous studies of CO CNe \citep{starrfield_2020_aa} and ONe CNe \citep{starrfield_2024_aa},
imply that the mass of the WD is growing toward the Chandrasekhar Limit and will either end as a Supernova Ia explosion or experience
Accretion Induced Collapse \citep{munari_1992_aa, ruiter_2019_aa, ruiter_2024_aa}.  
We use the Retention Efficiency (RE: (M$_{\rm acc}$ -  M$_{\rm ej}$)/M$_{\rm acc}$) to show this result.
Table \ref{tcrbtable2} gives the RE for all 15 simulations and the lowest value (0.56) is for the CO 50/50 simulation on the 1522 km WD.
In many other cases the value of the RE exceeds 0.9 which implies that the WDs are growing rapidly in mass.  Given that the mass of
the WD in TCrB is $\sim 1.35$ M$_\odot$ and the \.M necessary for the outburst to repeat on an 80 yr interval is 
$\sim 2 \times 10^{-8}$ M$_\odot$ yr$^{-1}$ it should take only a few million years to reach the above consequences.

As reported in \citet{starrfield_2020_aa, starrfield_2024_aa} for CNe,
our results in this paper for TCrB show that a significant amount of $^7$Be, which decays to $^7$Li, is produced by the TNR.  
The initial $^7$Li is destroyed by the TNR whereas $^7$Be is produced by the outburst and observed \citep{molaro_2022_aa, molaro_2023_aa}. 
Even though the distance to TCrB is $\ltsimeq 1$~kpc, the combination of low ejected mass and the mass fraction of $^7$Be for the largest amounts of $^7$Be produced in our simulations for TCrB, are too small for the decay of $^7$Be to $^7$Li to be detected by the INTErnational Gamma-Ray Astrophysics Laboratory (INTEGRAL). 

As in \citet{starrfield_2024_aa}, we take the ejected mass and mass fraction of $^7$Be and predict the 478 keV $\gamma$-ray emission from the decay of $^7$Be using \citet[][Eq. 7]{gehrz_1998_aa}. We find for the 1522 km WD CO 50/50 sequence a predicted value of $\sim 1.3 \times10^{-6}$ photons cm$^{-2}$ s$^{-1}$ and for the 1827 km WD CO 50/50 sequence a 
predicted value of $\sim 1.0 \times10^{-6}$ photons cm$^{-2}$ s$^{-1}$. These values are too small for a detection by INTEGRAL as derived by
\citet[][and references therein]{hernanz_2014_aa, kuulkers_2021_aa}.  All the other sequences predict smaller nuclear  $\gamma$-ray emission from TCrB.  
{However, it is possible that if the $^7$Be abundance is as large as found in the observations \citep{molaro_2023_aa},
then its decay could influence the decline in the CN or RN light curves.}

An important nucleus produced by both CO and ONe novae \citep{bekki_2024_aa} is $^{31}$P which is important for life on the Earth.  Our results show that it is also
produced on time scales characteristic of RNe such as TCrB whether the underlying WD is CO or ONe.  An important wavelength region to use
in studies of TCrB, therefore, is the infrared since infrared spectra of V1716 Sco when analyzed with CLOUDY \citep{ferland_2023_aa}, showed that phosphorus is highly enriched in the ejecta \citep{woodward_2024_aa}.

\section{Conclusions}
\label{conclude}
 We do not repeat the conclusions presented in either \citet{starrfield_2020_aa} for CO novae or in \citet{starrfield_2024_aa} for
 ONe novae but many of them are also appropriate for the work reported in this paper.

\begin{enumerate}

\item The oxygen-neon GR radius results in simulations that produce more explosive events resembling the observed outbursts of TCrB. However, our predicted peak ejecta velocities of $\sim$ 3600 km s$^{-1}$ are less than the $\sim$5000 km s$^{-1}$  reported for TCrB in the 1946 outburst by \citet{morgan_1947_aa} using photographic spectra.  Although those velocities seem reasonable, it will be important to determine the velocities using modern techniques. 

\item The measured IR abundances of the Red Giant in TCrB \citep{pavlenko_2020_aa, woodward_2020_aa} provide a foundation that allows us to determine the actual abundances of the material ejected from the WD, when the ejecta abundances are measured after the outburst.

\item  All the simulations predict more $^{13}$C is produced than $^{12}$C, and the  
CO 50/50 simulation on the 1522 km WD predicts an ejected amount of $^{13}$C that exceeds 5\% by mass fraction.  
 \citet{sneden_1975_aa} reported enriched $^{13}$C and $^{15}$N in the ejecta of Nova DQ Her(1934),
\citet{banerjee_2016_aa} reported $^{12}$C/$^{13}$C = 1.5 in V5668 Sgr,  \citet{joshi_2017_aa} reported $^{12}$C/$^{13}$C = 1.6 in
Nova  Oph 2017, and \citet{rudy_2024_aa} reported a value of $^{12}$C/$^{13}$C = 2 in V1391 Cas.  To our knowledge, however, there are no observations of CN ejecta that suggest that the abundance of  $^{13}$C exceeds that of $^{12}$C. Therefore, we suggest that the ejected material mixes with a large amount of material surrounding the WD and the binary system.

\item Recurrent novae such as TCrB produce $^7$Li in such amounts that if we factor in the number of outbursts, compared to typical CNe, they are also important contributors to the galactic amount of $^7$Li and maybe $^{26}$Al \citep{vasini_2024_aa}.

\item All our simulations result in the WD gaining mass at rates consistent with the mass accretion rate. 

\item Evolving the WD after the ejected matter has been removed from the simulations results in extremely high predicted
luminosities (L $>$$10^5$L$_\odot$) and effective temperatures (T$_{\rm eff}$$>$$2\times 10^6$K).  While two of the simulations result in evolution times that are close to those predicted by \citet{munari_2023_ab} for the second maximum, the other simulations bracket his predicted times.

\item Thus, TCrB, which is less than a kpc will become, for a short time, the brightest nova ever to be observed in X-rays. T CrB is an excellent candidate for detailed Swift and Chandra investigations.

\end{enumerate}
\medskip

 {\it Acknowledgements} 
 We are extremely grateful to U. Munari and S. Kenyon for their comments on an early draft which helped clarify the material in this manuscript.  We are also grateful to the anonymous referee for their comments which also helped improve this manuscript.  We acknowledge useful discussion and encouragement from A. C\'orsico, M. Darnley, E. Aydi, J. Jos\'e,  M. Hernanz, S. Kafka,  L. Izzo, P. Molaro, M. della Valle, and A. Shafter. This work was supported in part by the U.S. DOE under Contract No. DE-FG02- 97ER41041. SS and MB acknowledge partial support from a NASA Emerging Worlds grant to ASU (80NSSC22K0361) as well as support to SS from his ASU Regents' Professorship, 
This project has received funding [in part] from the European Union's Horizon Europe research and innovation program under grant agreement No. 101079231 (EXOHOST), and from UK Research and Innovation (UKRI) under the UK government Horizon Europe funding guarantee (grant number 10051045). WRH is supported by the U.S. Department of Energy, Office of Nuclear Physics, and CEW acknowledges support from NASA.   
This work has made use of data provided by Digital Access to a Sky Century @ Harvard (DASCH), which has been partially supported by NSF grants AST-0407380, AST-0909073, and AST-1313370. This research made use of the AAVSO Photometric All-Sky Survey (APASS), funded by the Robert Martin Ayers Sciences Fund and NSF AST-1412587.
{\bf Data Availability}
The data underlying this publication will be shared on reasonable request to the first author.

\bibliography{references_iliadis,starrfield_master}

\end{document}